\documentclass[12pt]{article}
\usepackage{amsmath,cite}
\usepackage{amssymb}
\usepackage[hidelinks]{hyperref}
\usepackage{cite}
\newcommand{\p}[1]{(\ref{#1})}
\newcommand{\be}{\begin{equation}}
\newcommand{\ee}{\end{equation}}
\newcommand{\bea}{\begin{eqnarray}}
\newcommand{\eea}{\end{eqnarray}}
\def\theequation{\arabic{section}.\arabic{equation}}
\begin{document}
\setcounter{page}0
\renewcommand{\thefootnote}{\fnsymbol{footnote}} 
\begin{titlepage}
\vskip .7in
\begin{center}
{\Large \bf  Cubic Vertices  for ${\cal N}=1$  Supersymmetric  Massless  Higher
Spin Fields in Various Dimensions
 } \vskip .7in 
 {\Large 
I.L. Buchbinder$^{a,b}$\footnote{e-mail: {\tt  joseph@tspu.edu.ru }}, 
V.A. Krykhtin$^a$\footnote{e-mail: {\tt  krykhtin@tspu.edu.ru }}, \\
M. Tsulaia$^c$\footnote{e-mail: {\tt  mirian.tsulaia@oist.jp }}, 
 D. Weissman$^c$\footnote{e-mail: {\tt  dorin.weissman@oist.jp}  }}
 \vskip .4in {$^a$ \it Department of Theoretical Physics, Tomsk State Pedagogical University, \\ 634041 Tomsk, Russia} \\
  \vskip .1in {$^b$ \it National Research  Tomsk State  University, \\ Lenin Av. 36, 634050 Tomsk, Russia} \\
\vskip .1in { $^c$ \it Okinawa Institute of Science and Technology, \\ 1919-1 Tancha, Onna-son, Okinawa 904-0495, Japan}\\
\vskip .8in
\begin{abstract}

Using the BRST approach to higher spin field theories we develop a generic technique for constructing the cubic interaction vertices for $N=1$ supersymmetric massless higher spin fields on four, six and ten dimensional flat backgrounds.  Such an approach allows formulation of the  equations for cubic vertices including bosonic and fermionic higher spin fields, and the problem of finding the vertices is reduced to finding the consistent solutions to these equations. As a realization of this procedure, we present the particular solutions for the vertices where the fields obey some off-shell constraints. It is shown that the supersymmetry imposes additional constraints on the vertices and singles out a particular subclass of the solutions. As a concrete application of the generic scheme, we consider supersymmetric Yang-Mills-like systems in four, six and ten dimensions where the higher spin fields transform under some internal symmetry group, as well as supergravity-like systems in the same dimensions.

\end{abstract}

\end{center}

\vfill

\end{titlepage}

\renewcommand{\thefootnote}{\arabic{footnote}}
\setcounter{footnote}0

\section{Introduction}

The problem of constructing interactions for  higher spin fields  has attracted much  attention for many years. Interest in this problem is due to certain possibilities for the development of new principles for constructing unified models of fundamental interactions, including quantum gravity, and phenomenological applications that are opening up in elementary particle physics and cosmology. Another fundamental principle that apparently should underlie the unified theory is related to supersymmetry. Therefore, the construction of a supersymmetric theory of higher-spin fields seems to be quite natural and relevant. 

Until now supersymmetric higher spin theories have been studied mainly in three and four dimensional flat and $AdS$ backgrounds 
(see 
\cite{Curtright:1979uz}--\cite{Kuzenko:1994dm} for earlier papers on the subject and \cite{Sezgin:2012ag}--\cite{Hutomo:2020wca}
for recent reviews),
whereas  considerations in higher dimensions have been relatively rare   \cite{Florakis:2014aaa}--\cite{Sorokin:2018djm}.
Nevertheless, higher dimensional supersymmetric higher spin theories are interesting for several reasons.
Firstly, they might prove to  be helpful for further investigations of a connection between
higher spin  and string theories, since the latter also lives in higher dimensions.
Secondly, 
higher dimensional higher spin  theories can open some new possibilities
for building lower dimensional supersymmetric models,
as in supergravity,
where the lower dimensional models can be obtained from relatively simpler
higher dimensional ones via various kinds of compactifications and dimensional reductions.

Recently a particular 
  class of free Lagrangians for $N=1$ supersymmetric massless
higher spin fields was obtained  \cite{Sorokin:2018djm},\cite{Fotopoulos:2008ka} for  $D=3,4,6$ and $10$ dimensional flat backgrounds. 
Although this construction bears a certain similarity with the supersymmetric open string field theory \cite{Kazama:1986cy}, 
 it turns out that for massless higher spin fields
one can build  finite dimensional supermultiplets with the supersymmetry algebra
being closed  on-shell  both in the bosonic (NS) and in the fermionic (R) sectors.
To be more specific, the model obtained   in \cite{Sorokin:2018djm} is 
a supersymmetrization of 
free Lagrangians for certain reducible representations of
the Poincar\'e
group and  the specific structure of these representations 
is singled out by the requirement of $N=1$ supersymmetry.
The fermionic sector (so-called  ``triplet''  \cite{Francia:2002pt}--\cite{Agugliaro:2016ngl})
contains  physical and auxiliary fields, which are totally symmetric with respect
to their indices.\footnote{A frame-like formulation
of triplets was obtained in \cite{Sorokin:2008tf}.}
The physical field and  a part of the auxiliary fields 
in the bosonic sector have indices of two types:
$n$ indices of one type and  one index of another type.
The indices of the first type are totally symmetric among
each other, while there is no symmetry between them and the index of the second type.
The other auxiliary fields are totally symmetric, i.e., they have indices only of the first type.
This is the simplest example of so--called
generalized triplets \cite{Sagnotti:2003qa} which describes reducible representations
of the Poincar\'e group for fields with mixed symmetries.
In both the fermionic and bosonic sectors the auxiliary fields
are eliminated
via their
own equations of motion and/or  after the complete gauge fixing so the system
describes on-shell only physical polarizations.
To summarize, in the fermionic sector the physical fields are described by the rank $n$
spin tensor which  
contains the fields with spins $n+1/2,n-1/2,...,1/2$
The bosonic sector contains the physical fields
described by Young tableaux with two rows. These Young tableaux are of the type
$(n,1)$ and $(n+1,0)$.

As mentioned above, the triplets are reducible representations of the Poincar\'e group
and unlike the so-called Fronsdal fields \cite{Fronsdal:1978rb}--\cite{Fang:1978wz}, each triplet  contains more than one physical  field. These
fields correspond  to single, double, etc. traces of the tensor/spin-tensor field
of rank $n$. On the other hand, the question of whether the corresponding
supermultiplets are reducible or not in the sense of representations of SUSY algebra
has different answers depending on the space-time dimensions, as they can be either reducible or irreducible. 
This can be easily seen for the example of the lower spin fields.
The lowest spin case which corresponds to $n=0$, describes the $N=1$ supersymmetric Maxwell theory
in $D=4,6$ and $10$ and is in some sense degenerate. The next simplest case,
with $n=1$ corresponds to linearized $N=1$ supergravity theories.
The Lagrangians of \cite{Sorokin:2018djm} describe   irreducible
supermultiplets for $D=10$  and reducible supermultiplets for $D=4$ and $D=6$. Proceeding further, one can show
that the  situation 
for the higher spin fields  is the same as for the case of  linearized supergravity multiplets.

It is a natural next step to study a possibility
for cubic interactions for the free systems described above.
The problem of construction and of the further study of
cubic vertices for massless and massive higher spin fields
has been  attracting  considerable interest \cite{Metsaev:2005ar}--
\cite{Sleight:2017pcz} (see also \cite{Bengtsson:1986kh}--\cite{Koh:1986vg} for earlier work
and \cite{Vasiliev:2001wa}-- \cite{Boulanger:2011qt}
for supersymmetric and non-supersymmetric
cubic interactions on $AdS$ backgrounds in the frame--like approach).
Although these studies were mainly devoted to non-supersymmetric theories, several
interesting results appeared recently  for supersymmetric cubic vertices in four dimensions.
In particular, in \cite{Metsaev:2019dqt}--
\cite{Metsaev:2019aig} cubic  vertices for $N=1$ superfields and for extended
$N$ were
obtained in the light-cone approach. In the papers
\cite{Buchbinder:2017nuc}-- \cite{Buchbinder:2018nkp} 
the interactions between conserved higher spin supercurrents
and chiral superfields, as well as interactions between supersymmetric sigma-models 
and higher spin superfields on flat and $AdS$ backgrounds were constructed.
The cubic interactions for supersymmetric systems were also recently constructed in
\cite{Bonora:2020aqp}-- \cite{Khabarov:2020deh}.

In the present paper we extend, at least partially, the results of \cite{Sorokin:2018djm}
by including cubic interaction vertices into consideration.
By  ``partially'' we mean the following: 
for the purpose of simplifying the computations, we shall partially gauge fix 
the free Lagrangians, so the fields contain only physical transverse
components. As the second step we perform nonlinear deformations
of these Lagrangians by including cubic interaction vertices.
It turns out however, that because of this gauge fixing, a further requirement of
the invariance under $N=1$ supersymmetry transformations
puts the fields completely on shell. 
These completely on-shell vertices can be promoted back to
the off-shell ones by including  all auxiliary fields into the free
and interacting
Lagrangians
i.e., by considering the system given in \cite{Sorokin:2018djm} without gauge fixing
and then repeating the procedure described above.\footnote{
Usually in the context of supersymmetric theories the words
``on-shell'' and ``off-shell'' indicate whether  the supersymmetry algebra is closed after taking into account the equations of motion or not
 (see e.g. \cite{Buchbinder:1995uq}).
 The $N=1$ systems constructed in \cite{Sorokin:2018djm} are formulated in terms of component
 fields, i.e., they are ``on-shell''.
  Here  ``completely on-shell'' means that
 we use the field equations
  in order to have the cubic vertices transformed into each other under the supersymmetry 
  transformations.} 
  The ``physical'' part of these vertices, which does not contain any auxiliary fields will coincide
with the ones obtained in the present paper.
Here we shall present the defining equations for these vertices and leave
the detailed analysis for a separate publication.

 The paper is organized as follows:

In section \ref{QL} we give a brief description of the free supersymmetric systems  
for which we are going to build cubic interactions. To this end, we use the BRST approach.\footnote{Originally used for a description
of totally symmetric massless and massive reducible representations of the Poincar\'e group
\cite{Ouvry:1986dv}--\cite{Bengtsson:1986ys}, this approach was then generalised
for description of Poincar\'e and $AdS_D$ groups, see e.g.  \cite{Pashnev:1998ti}--\cite{Alkalaev:2018bqe}. }
which yields the free Lagrangians given in \cite{Sorokin:2018djm} with no off-shell constraints neither on the fields under consideration
nor on the parameter of gauge transformations.
We then gauge fix the Lagrangians, so that they contain only physical components,
and the fields  and parameters of gauge transformations obey certain off--shell constraints
\footnote{This is called a constrained formulation in \cite{Buchbinder:2007ak}
whereas the formulation where no off-shell constraints are imposed is called unconstrained one
\cite{Buchbinder:2008ss}.}

Section \ref{CI}, where we describe the cubic interactions,
contains two subsections.
In subsection \ref{subbbb}
we collect the expressions for the vertices for three bosonic
higher spin fields both in an unconstrained and in  a gauge fixed form
\cite{Fotopoulos:2010ay}, \cite{Metsaev:2007rn}.
These vertices correspond to the purely  bosonic part of cubic interactions
of the systems under consideration.
The equations that determine 
cubic vertices for two fermionic and one bosonic higher spin fields are given
 in  subsection \ref{subffb}.

In section \ref{ssym} we describe the higher spin generalization
of the $N=1$ super Yang-Mills theory.
First we derive the corresponding cubic vertices and the Lagrangians,
which describe the cubic interactions between reducible 
massless representations of the Poincar\'e group.
These vertices are  a covariant
form of the analogous vertices obtained in the light-front
formalism in \cite{Metsaev:2007rn}.
Then we present the $N=1$ supersymmetry transformation for these systems.
As mentioned above, the requirement of the supersymmetry puts the fields on shell, because of the form
of the gauge fixing.
A somewhat degenerate case of $N=1$ super Yang-Mills theory
which illustrates how the whole system can be promoted  to an off-shell description
is given separately.

An analogous consideration for the higher spin generalisation 
of the four dimensional $N=1$ supergravity is given in section
\ref{ssugra}.

The last section contains our conclusions and the summary of results.

Some lengthy equations and useful identities for gamma matrices and for linearized gravity are collected in the appendices.

\section{Free Lagrangians} \label{QL} \setcounter{equation}0 

In this section we shall
briefly describe free Lagrangians for bosonic and fermionic massless higher spin fields,
whose cubic deformations and supersymmetrizations we are going to consider.

As we mentioned  in the introduction,
in the fermionic $F$ sector we have $n$ totally symmetric  spin-tensor fields, both physical and auxiliary.
In the bosonic sector $B$   the physical field contains
 $n$ indices which are symmetric among each other and one index which has no symmetry with the other ones.
 The auxiliary fields in the bosonic sector are either totally symmetric or have   mixed symmetry.

In order to derive the corresponding Lagrangians, we introduce
commuting oscillators $\alpha^{\mu, \pm}_m$,   anticommuting ghosts $c_m^{\pm}$, $c_0$
and antighosts $b_m^{\pm}$, $b_0$, where $m=1,2$ in the $B$ sector and 
$m=1$ in the $F$ sector.
These  oscillators obey the following (anti)commutation relations
\begin{equation}\label{B4}
[\alpha^\mu_m, \alpha^{\nu, +}_n ] =  \eta^{\mu \nu}\delta_{mn},
\quad \{ c^{+}_m, b_n \} = \{ c_m, b^{+}_n \} = 
 \delta_{mn}\,, \quad \{ c_0 , b_0 \}=1,
\end{equation}
The ghost number of $c^{\pm}_m$ and $c_0$ is $+1$, the ghost number of  $b^{\pm}_m$ and $b_0$ is consequently $-1$
and the ghost number of  $\alpha^{\mu, \pm}_m$  is zero.

  The Fock vacua in the $B$ and in $F$ sectors are
defined as, respectively
\be\label{vacuumns-1}
\alpha^\mu_m|0_{B} \rangle =  
c_m|0_{B} \rangle = b_m|0_{B} \rangle = b_0 |0_{B} \rangle=0, \qquad m=1,2.
\ee
\be\label{vacuumr-1}
\alpha^\mu_1|0_F \rangle =  
c_1|0_F \rangle = b_1|0_F \rangle = b_0 |0_{F} \rangle
=0.
\ee
Higher spin functionals  either in the $B$ or $F$ sector,  are expanded in terms of the creation operators  and the components of this expansion are higher spin fields (physical and auxiliary).

Let us now introduce differential operators. In the $B$ sector we have
\be\label{l}
l_0 = p\cdot p, \qquad l_m = p \cdot \alpha_m, \qquad l^{+}_m = p \cdot \alpha^{+}_m,
\ee
where $p_\mu = - i \partial_\mu$ when acting to the right. The symbol
`dot' means
 $A \cdot B= \eta_{\mu\nu} A^\mu B^\nu$ and $\partial  A$  denotes a symmetrized derivative. For example if $A$  is a vector, then $\partial  A \equiv \partial_\mu A_\nu + \partial_\nu A_\mu$. 
Obviously $l_0$ is the d'Alembertian, $l_m$ being divergence operators with respect to the indices
contracted with $\alpha^{\mu,+}_m$ oscillators and $l_m^+$ being derivatives symmetrized
with
the indices
contracted with $\alpha^{\mu,+}_m$ oscillators. Alternatively, one can work in a momentum representation, without
realizing $p_\mu$ as a differential operator.

In the $F$ sector we have
\be \label{2222}
l_1 = p \cdot \alpha_1, \qquad l^{+}_1 = p \cdot \alpha^{+}_1,
\quad
g_0 = p \cdot \gamma, 
\ee 
where $\gamma_\mu$ are gamma-matrices and the operator  $g_0$ being the Dirac operator.

Having defined all necessary operators we can write 
 a free   Lagrangian  for bosonic fields as
\be \label{falphal}
{ \cal L}_{B,\text{free}}= \int d c_0 \langle {\Phi}_{B} | Q_{B} | \Phi_{B} \rangle.
\ee
with the corresponding nilpotent BRST charge 
\be \label{BRSTNS} Q_{B} \ = \ c_0 \,
l_0 \ + \ \tilde Q_{B} \ - \ M_{B} \, b_0 \ , 
\ee 
where
\be
 \label{TQ} \tilde Q_{B}\ = \  \; \;
\sum_{m=1,2}(c^+_m \, l_m   + c_m \, l^+_m) ,  \\
\qquad 
M_{B}\ = \  
\;  \,  \, \sum_{m=1,2} c^+_m\, c_m  ,
 \ ,
\ee
The Lagrangian \p{falphal}  is invariant under the gauge transformations
\be \label{GAUGENS}
\delta |\Phi_{B} \rangle = Q_{B} | \Lambda_{B} \rangle
\ee
due to the nilpotency of the BRST charge \p{BRSTNS} in any space-time dimension $D$.
The ghost number of the field $| \Phi_{B} \rangle$ is fixed to be zero and consequently
the ghost number of the parameter of gauge transformations $| \Lambda_{B} \rangle$
is equal to $-1$.

Further, in order to establish $N=1$ supersymmetry, one requires that the component
of the higher spin functional  $| \Phi_{B} \rangle$, which does not depend on ghosts/antighosts
and describes the physical field,  depends  on the oscillator  $\alpha_2^{+\nu} $ only linearly.
The explicit form of $| \Phi_{B} \rangle$ and $| \Lambda_{B} \rangle$  can be completely
fixed
and is given in \cite{Sorokin:2018djm}.

In the following we  shall work with the gauge fixed form
of the Lagrangian \p{falphal} by imposing  off-shell  conditions
$| \Phi_{B} \rangle$
\be \label{GFFF}
l_m | \Phi_{B} \rangle=0, \quad m=1,2.
\ee
As a result  the higher spin functional contains only a physical field
\be
|\Phi_{B} \rangle \equiv |\phi \rangle  =
\frac{1}{n!}\ \phi_{\mu_1\mu_2\ldots
\mu_n,\nu}(x)\, \alpha^{\nu+}_2  \alpha^{\mu_1+}_1 \, \alpha^{\mu_2 +}_1 \ldots
\alpha^{\mu_n +}_1   \, |0_{B} \rangle,
\ee
all ghost dependence being gauged away.
The physical field obeys off-shell transversality conditions, i.e.,
we are essentially dealing with only physical components.

The  Lagrangian \p{falphal} and the gauge fixing conditions \p{GFFF}  
are still invariant under the gauge transformations \p{GAUGENS}
with  the parameter of gauge transformations
\bea \label{NStrip1}
&&|\Lambda_{B} \rangle = b^+_1 |\lambda \rangle + b^+_2 |\rho \rangle = \\ \nonumber
&&  =  \frac{i b^+_1}{(n-1)!} \lambda_{\nu, \mu_1\mu_2\, ...\,
\mu_{n-1} }(x)\, \alpha^{\nu+}_2\, \alpha^{\mu_1+}_1 \, \alpha^{\mu_2 +}_1 \, ...\,
\alpha^{\mu_{n-1} +}_1   \, |0_{B} \rangle + \\ \nonumber
 &&+ \frac{i b^+_2}{n!} \rho_{\mu_1\mu_2\, ...\,
\mu_n}(x)\, \alpha^{\mu_1+}_1 \, \alpha^{ \mu_2+}_1 \, ..\,
\alpha^{ \mu_n +}_1 \, |0_{B} \rangle \\ \nonumber 
\eea
being restricted as
\be
l_0|\Lambda_B \rangle = l_m |  \Lambda_B\rangle =0, \quad m=1,2
\ee
The free Lagrangian for the fermionic triplet is
\bea \label{ftr-free}
{ \cal L}_{F,\text{free}}&=&\frac{1}{{\sqrt 2}}\,\,{}_a\langle \Phi_{F,1}|(g_0)^a{}_b|\Phi_{F,1}\rangle^b +
 {}_a\langle \Phi_{F,2}|\tilde Q_{F}|\Phi_{F,1}\rangle^a + \\ \nonumber
&+& {}_a\langle \Phi_{F,1}|\tilde Q_{F}|\Phi_{F,2}\rangle^a +
\sqrt{2}\,\,{}_a\langle \Phi_{F,2}|M_F (g_0)^a{}_b|\Phi_{F,2}\rangle^b\,.
\eea
where
\be
 \label{TQ-R}
\tilde Q_{F}=
c^+_1 \, l_1   + c_1 \, l^+_1,  \qquad
M_{F} =   
\;  \,  \,  c^+_1\, c_1,
\ee
 The Lagrangian contains two fields, each of them being a series of expansion in terms of the creation operators
 $\alpha^{\nu, +}_1$, $c^{+}_1$, and  $b^{+}_1$. 
 The field $|\Phi_{F,1}\rangle^b$ contains  physical and  auxiliary fields, while the field
 $|\Phi_{F,2}\rangle^a$ is purely auxiliary.
 The Lagrangian \p{ftr-free} is invariant under the gauge transformations
 \begin{eqnarray} \label{GTR1}
&& \delta\, |\Phi_{F,1}\rangle^a \ = \ \tilde Q_F |\Lambda_{F,1} \rangle^a
\ +\ \sqrt{2} \,M_F \, (g_0)^a{}_b \, |\Lambda_{F,2} \rangle^b \ , \label{GT1}
\nonumber \\
&& \delta \,|\Phi_{F,2}\rangle^a \ = \ - \frac{1}{\sqrt{2}}(g_0)^a{}_b \, |\Lambda_{F,1} \rangle^b \ - \ \tilde
Q_F \, |\Lambda_{F,2} \rangle^a \  
\end{eqnarray}
with unconstrained parameters $|\Lambda_{F,1} \rangle^a$ and $|\Lambda_{F,2} \rangle^a$. 
The fields $|\Phi_{F,1}\rangle^a$ and $|\Phi_{F,2}\rangle^a$ have ghost numbers
$0$ and $-1$ respectively. The parameters of gauge transformations
$|\Lambda_{F,1} \rangle^a$ and $|\Lambda_{F,2} \rangle^a$
have ghost numbers $-1$ and $-2$ respectively.
Similarly to the bosonic sector, 
one can consider
only a physical  field
\be\label{Rtrip}
|\Phi_F\rangle^a  
 \equiv |\Psi\rangle^a  =
\frac{1}{n!}\ \Psi^a_{\mu_1\mu_2\, ...\,
\mu_n}(x)\, \alpha^{\mu_1+}_1 \, \alpha^{\mu_2 +}_1 \, ...\,
\alpha^{\mu_n +}_1\, |0_F \rangle. 
\ee
 In this gauge the field 
$|\Phi_F\rangle^a $ does not depend on 
ghost/antighost variables  and  obeys the off-shell transversality condition
\be \label{GENTR}
l_1 | \Phi_F \rangle^a =0.
\ee
The free Lagrangian in the fermionic sector is therefore 
\be \label{f-free}
{ \cal L}_{F.free}= {}_a\langle  \Phi_F| (g_0)^a{}_b | \Phi_f \rangle^b.
\ee
 The off-shell constraints \p{GENTR} and the Lagrangian \p{f-free}
 are invariant
 under the transformations
 \be \label{GTR1-1-1}
 \delta\, |\Phi_F\rangle^a \ = \ \tilde Q_F |\Lambda_F \rangle^a
 \ee
  provided  the parameter of gauge transformations
  \begin{equation}
|\Lambda_F \rangle^a \ =\ b^+_1 |\tilde \lambda \rangle^a\,=
\frac{ib^+_1}{(n-1)!}\
\tilde \lambda^a_{\mu_1\mu_2...\mu_{n-1}}(x) \, \alpha^{+\mu_1}_1 \,
\alpha^{+\mu_2}_1 ... \alpha^{+\mu_{n-1}}_1\, |0_F\rangle \ ,  
\end{equation}
  is constrained as
\be \label{const-gauge}
(g_0)^a{}_b |   \Lambda_F \rangle^b= l_1 | \Lambda_F \rangle^a
=0.
\ee
Finally, let us note that all representations of the Poincar\'e
group that we discussed in this section  are reducible since no  (gamma)tracelesness condition has been imposed at any stage.

\section{Cubic interactions} \label{CI}\setcounter{equation}0
\subsection{Three bosons}\label{subbbb}
Below we shall follow the approach of
\cite{Buchbinder:2006eq}  
for the construction of off-shell cubic interaction vertices between three bosonic  higher spin fields.
This method is a modification of the construction developed in open string field theory
\cite{Neveu:1986mv} -- \cite{Gross:1986ia} to the case of higher spin fields and it can be applied
for either massless or massive fields, both on flat and $(A)dS_D$ backgrounds.

Let us take three copies
of the Fock spaces introduced in section \ref{QL}.
All oscillators  get an extra index $i=1,2,3$
and the nonzero commutation relations are only between oscillators, which belong to the same Fock space 
\begin{equation}\label{B4-c}
[\alpha_{\mu,m}^{(i)}, \alpha_{\nu,n}^{(j), +} ] =  \delta^{ij}\delta_{mn}\eta_{\mu \nu},
\ee
\be
\{ c^{(i), +}_m, b^{(j)}_n \} = \{ c^{(i)}_m, b^{(j),+}_n \} 
= 
\{ c_{0,m}^{(i)} , b_{0,n}^{(j)} \} =
 \delta^{ij}\delta_{mn}\,,
\end{equation}

The operator $p_\mu^{(i)}$ 
 corresponds to the momentum in the $i$-th Fock space.
In a coordinate representation
the expression $p_\mu^{(i)}=-i \partial_\mu^{(i)} $ is a derivative acting on the fields in $i$-th Fock space.
The momentum operators obey the constraint
\be
p_\mu^{(1)} + p_\mu^{(2)} + p_\mu^{(3)}=0.
\ee
Finally, we also allow the higher spin functionals to carry some internal  symmetry indices denoted as  $A,B,C$.

Next, we consider the Lagrangian
\bea \label{LNSINT}
{ \cal L}_{B,\text{int}} &=& \sum_{i=1}^3 \int dc_0^{(i)} 
{}^A\langle \Phi_{B}^{(i)} |Q_{B}^{(i)}| \Phi_{B}^{(i)}
\rangle_A + \\ \nonumber 
&+&g \left ( \int dc_0^{(1)} dc_0^{(2)} dc_0^{(3)} {}^A\langle \Phi_{B}^{(1)}| {}^B\langle \Phi_{B}^{(2)}| {}^C\langle \Phi_{B}^{(3)}| |V \rangle_{ABC} + h.c.
\right )
\eea
and modified gauge transformations
\bea \label{GTNSINT}
\delta | \Phi_{B}^{(1)} \rangle_A & = & Q_{B}^{(1)} |\Lambda_{B}^{(1)} \rangle_A
-  \\ \nonumber 
&-& g \int dc_0^{(2)} dc_0^{(3)} 
\left ( ({}^B\langle \Phi_{B}^{(2)}| {}^C\langle \Lambda^{(3)}_{B}|
+ {}^C\langle \Phi_{B}^{(3)}| {}^B\langle \Lambda_{B}^{(2)}|) | V \rangle_{ABC} \right )
\eea
 where $g$ is a coupling constant.
 The transformations for the fields $ | \Phi_{B}^{(2)} \rangle_A$  and
$ |\Phi_{B}^{(3)} \rangle_A$ are obtained from \p{GTNSINT}
via cyclic permutations. 

Below, we will consider two types of vertices. In one type of vertices,
to which we refer as ``gravity-like'' vertices, the internal indices are absent.
The other type of vertices, referred to as ``Yang-Mills-like'', have the form
\be |V\rangle_{ABC}=f_{ABC} |V\rangle \ee
for some totally antisymmetric structure constants \(f_{ABC}\).

In both cases the invariance of the Lagrangian \p{LNSINT} under the transformations \p{GTNSINT} in the zeroth order in $g$ 
is maintained due to the nilpotency of the BRST charges
in each Fock space
\be
(Q_{B}^{(1)})^2 = (Q_{B}^{(2)})^2 = (Q_{B}^{(3)})^2=0.
\ee
The invariance at the first order in $g$
implies  that the cubic vertex is BRST invariant
\be \label{NSBRSTV}
(Q_{B}^{(1)} + Q_{B}^{(2)}+ Q_{B}^{(3)})  | V \rangle=0
\ee
The condition \p{NSBRSTV} also guarantees that 
the group structure of the gauge transformations is preserved at the first order in $g$. 

The cubic vertex has a general structure
\be \label{V3Bosons}
|V \rangle = V \,\, c_0^{(1)} c_0^{(2)} c_0^{(3)} \,\, 
| 0_B^{(1)} \rangle \otimes
|0_B^{(2)} \rangle \otimes | 0_B^{(3)} \rangle
\ee
where an unknown function $V$ can depend on  $p_\mu^{(i)}, 
\alpha^{(i),+}_\mu,  c^{(i),+}, b^{(i),+}, b_0^{(i)+}$.
Apart from the condition of BRST invariance \p{NSBRSTV},
the function $V$ is required to be 
 Lorentz invariant and to have zero ghost number.

It can be verified  by direct computations \cite{Fotopoulos:2010ay}, \cite{Metsaev:2012uy}
that the following expressions are BRST invariant and therefore any function of them is a solution
of 
\p{NSBRSTV}
\be \label{sbv-1-x}
{\cal K}^{(i)}_m = 
(p^{(i+1)}- p^{(i+2)}) \cdot \alpha^{(i),+}_m +
(b_0^{(i+1)}-b_0^{(i+2)}) \, c^{(i),+}_m
\ee

\be \label{sbv-10}
{\cal O}^{(i,i)}_{mn} =  \alpha^{(i),+}_m \cdot \alpha^{(i),+}_n
+ c^{(i),+}_mb^{(i),+}_n + c^{(i),+}_nb^{(i),+}_m
\ee

\be \label{sbv-3}
{\cal Z}_{mnp}= {\cal Q}^{(1,2)}_{mn} {\cal K}^{(3)}_p +
{\cal Q}^{(2,3)}_{np} {\cal K}^{(1)}_m +
{\cal Q}^{(3,1)}_{pm} {\cal K}^{(2)}_n
\ee
where
\be \label{sbv-7}
{\cal Q}^{(i,i+1)}_{mn} = \alpha^{(i),+}_m \cdot \alpha^{(i+1),+}_n
+ \frac{1}{2} b^{(i),+}_m c^{(i+1),+}_n + \frac{1}{2} b^{(i+1),+}_n c^{(i),+}_m
\ee
After the gauge fixing described in section \ref{QL}
the Lagrangian \p{LNSINT} simplifies to
\be\label{L-bbb-gf}
{\cal L}_{\text{int}} = \sum_{i=1,2,3}
{}^A\langle \phi^{(i)} |l_{0}^{(i)}| \phi_{B}^{(3)}
\rangle_A + 
\left (  {}^C\langle \phi^{(1)} | \,\,
{}^A\langle \phi^{(2)}| \,\,
{}^B\langle \phi^{(3)}|  
| { V} \rangle_{ABC}
+ h.c \right )
\ee
and describes cubic interactions between  three massless higher spin fields
without their auxiliary components.
As we shall see below, further requirement of $N=1$ supersymmetry 
will single out some particular subclasses  of the cubic vertices.

\subsection{Two fermions and one boson} \label{subffb}

Cubic interactions   between two fermionic and one bosonic higher spin fields
can be treated in the BRST approach in a similar way. However, there is 
one important difference, which makes the present case technically more complicated. 
This difference shows up already at the level of the free Lagrangians:
because of the absence  of the ghost $c_0$
 (see 
section \ref{QL} and 
\cite{Sorokin:2018djm}, \cite{Sagnotti:2003qa}   for details) 
the free Lagrangian for the fermionic triplets contains two different operators: the Dirac operator
$g_0$ and operator \p{TQ-R}, instead of only one BRST charge \p{BRSTNS}
present in the Lagrangians for free  bosonic triplets.

Making a cubic deformation of the free Lagrangian
we get
\bea\label{L-ffb-gen}
{\cal L}_{\text{int}} &=& \sum_{i=1,2} (\frac{1}{\sqrt{2}}\,\,{}_a^A\langle \Phi_{F,1}^{(i)}|{(g_0^{(i)})^a{}_b}
|\Phi_{F,1}^{(i)}\rangle^b_A +
 {}_a^A\langle \Phi_{F,2}^{(i)}|\tilde Q_{F}^{(i)}|\Phi_{F,1}^{(i)}\rangle^a_A + \\ \nonumber
 &+& {}_a^A\langle \Phi_{F,1}^{(i)}|\tilde Q_{F}^{(i)}|\Phi_{F,2}^{(i)}\rangle^a_A +
\sqrt{2}\,\,{}_a^A\langle \Phi_{F,2}^{(i)}|M_F^{(i)} (g_0^{(i)})^a{}_b|\Phi_{F,2}^{(i)}\rangle^b_A ) \\ \nonumber
&+&  \int dc_0^{(3)} 
{}^A\langle \Phi_{B}^{(3)} |Q_{B}^{(3)}| \Phi_{B}^{(3)}
\rangle_A + \\ \nonumber
&+& g\sum_{m,n=1,2} \int dc_0^{(3)} 
\left (  {}^C\langle \Phi_{B}^{(3)} | \,\,
{}_a^A\langle \Phi_{F,m}^{(1)}| \,\,
{}_b^B\langle \Phi_{F,n}^{(2)}|  
| {\cal V}_{mn} \rangle^{ab}_{ABC}
+ h.c \right ).
\eea
The structure of the cubic vertex  
\be
| {\cal V} \rangle^{ab}_{ABC} = {\cal V}^{ab}_{ABC}\,\, |0_F^{(1)} \rangle \otimes
|0_F^{(2)} \rangle \otimes c_0^{(3)} | 0_B^{(3)} \rangle.
\ee
and  the fact that the  Lagrangian and the higher spin functionals have the ghost number zero, 
imply that the unknown function ${\cal V}^{ab}_{ABC}$ has ghost number zero as well.
The requirement of the invariance of the Lagrangian
\p{L-ffb-gen} under the nonlinear gauge transformations 
\bea \label{BFFG-1-gen} 
 \delta | \Phi_{B}^{(3)} \rangle_C &= & 
Q_{B}^{(3)} |\Lambda_{B}^{(3)} \rangle_C + \\ \nonumber
&+&g\sum_{m,n=1,2}  ( {}_a^A\langle { \Phi}_{F,m}^{(1)}| \,\,
{}_b^B\langle  { \Lambda_{F,n}}^{(2)}|  
|{\cal W}_{3,mn}^{1,2} \rangle^{ab}_{ABC}
+ \\ \nonumber
&+& {}_a^A\langle \Phi_{F,m}^{(2)}|  \,\, 
{}_b^B\langle { \Lambda}_{F,n}^{(1)}|  \
|{\cal W}_{3,mn}^{2,1} \rangle^{ab}_{ABC} )
\eea
\bea \label{BFFG-2-1-gen}
 \delta | \Phi_{F,1}^{(i)} \rangle^a_C &= & 
{\tilde Q}_{F}^{(i)} |{ \Lambda_{F,1}}^{(i)} \rangle^a_C +
 \sqrt{2} \,M_F^{(i)}  \, (g_0^{(i)})^a{}_b \, |\Lambda_{F,2}^{(i)}  \rangle^b_C\\ \nonumber
 &+&g  \sum_{m=1,2}\int dc_0^{(3)}(
 {}_b^A\langle \Phi_{F,m}^{(3-i)}|\,\,
{}^B\langle  \Lambda_{B}^{(3)}|   | {\cal W}_{i,1m}^{3-i,3} \rangle^{ab}_{ABC}
+ \\ \nonumber
&+&{ {}^A\langle \Phi_{B}^{(3)}|\,\, 
{}_b^B \langle \Lambda}_{F,m}^{(3-i)}| |
{\cal W}^{3,3-i}_{i,1m} \rangle^{ab}_{ABC} 
) 
\eea
\bea \label{BFFG-2-2-gen}
 \delta | \Phi_{F,2}^{(i)} \rangle^a_C &= & -
\frac{1}{\sqrt{2}} (g_0^{(i)})^a{}_b \, |\Lambda_{F,1}^{(i)} \rangle^b_C \ - \ \tilde
Q_F^{(i)} \, |\Lambda_{F,2}^{(i)} \rangle^a_C \\ \nonumber
 &+&g  \sum_{m=1,2}\int dc_0^{(3)}(
 {}_b^A\langle \Phi_{F,m}^{(3-i)}|\,\,
{}^B\langle  \Lambda_{B}^{(3)}|   | {\cal W}_{i,2m}^{3-i,3} \rangle^{ab}_{ABC}
+ \\ \nonumber
&+&{ {}^A\langle \Phi_{B}^{(3)}|\,\, 
{}_b^B \langle \Lambda}_{F,m}^{(3-i)}| |
{\cal W}^{3,3-i}_{i,2m} \rangle^{ab}_{ABC} 
) 
\eea
leads to equations for unknown vertices
$| {\cal V}_{mn} \rangle^{ab}_{ABC}$
and $| {\cal W}^{ij}_{k,mn} \rangle^{ab}_{ABC}$ which are given in
\p{eq-unf-1}--\p{eq-unf-12}.

One can however consider a simpler problem, where the 
higher spin functionals are  gauge fixed,
as  was discussed in section \ref{QL}. 
At the cubic level this simply means 
considering only physical (ghost independent) components in \p{L-ffb-gen}
and integrating out of the ghost zero mode.
Then the corresponding cubic Lagrangian has the form
\bea \label{int.}
{\cal L}_{\text{int}} 
&=&
\sum_{i=1}^2 
{}_a^A\langle \Psi^{(i)}|(g_0^{(i)})^a{}_b|\Psi^{(a)}\rangle^b _A + 
{}^A\langle \phi^{(3)}|
l_{0}^{(3)} | \phi^{(3)} \rangle _A + \\ \nonumber 
&&+ g\left (  \,\,    {}^C\langle \phi^{(3)} | \,\,
{}_a^A\langle \Psi^{(1)}| \,\,
{}_b^B\langle \Psi^{(2)}|  | {\cal V} \rangle^{ab}_{ABC}
+ h.c \right ).
\eea
The invariance under nonlinear gauge transformations 
\bea \label{BFFG-1} 
 \delta | \phi^{(3)} \rangle_C &= & 
\tilde Q_{B}^{(3)} |\Lambda_{B}^{(3)} \rangle_C + \\ \nonumber
&+& g ( {}_a^A\langle { \Psi}^{(1)}| \,\,
{}_b^B\langle  { \Lambda_{F}}^{(2)}|  
|{\cal W}_3^{1,2} \rangle^{ab}_{ABC}
+
{}_a^A\langle \Psi^{(2)}|  \,\, 
{}_b^B\langle { \Lambda}_F^{(1)}|  \
|{\cal W}_{3}^{2,1} \rangle^{ab}_{ABC} ),
\eea
\bea \label{BFFG-2-1}
 \delta | \Psi^{(1)} \rangle^a_C &= & 
{\tilde Q}_{F}^{(1)} |{ \Lambda_{F}}^{(1)} \rangle^a_C +  \\ \nonumber
 &+&g   (
 {}_b^A\langle \Psi^{(2)}|\,\,
{}^B\langle  \Lambda_{B}^{(3)}|   | {\cal W}_{1}^{2,3} \rangle^{ab}_{ABC}
+
{ {}^A\langle \phi^{(3)}|\,\, 
{}_b^B \langle \Lambda}_F^{(2)}| |
{\cal W}^{3,2}_1 \rangle^{ab}_{ABC} 
),
\eea
\bea \label{BFFG-2-2}
 \delta | \Psi^{(2)} \rangle^a_C &= & 
{\tilde Q}_{F}^{(2)} |{ \Lambda_{F}}^{(2)} \rangle^a_C + \\ \nonumber
 &+&g    ({}^A\langle \phi^{(3)}| \,\,
{}_b^B\langle  \Lambda_{F}^{(1)}|   | {\cal W}_{2}^{3,1} \rangle^{ab}_{ABC}
+
{}_b^A\langle \Psi^{(1)} |\,\,
{}^B\langle 
{ \Lambda}_{B}^{(3)}| |
{\cal W}_2^{1,3} \rangle^{ab}_{ABC} 
),
\eea
implies the following conditions on the vertices
\be \label{xy-1}
{}^B\langle \Lambda_B^{(3)}|{}^C_a \langle \Psi^{(1)} | {}^A_c\langle \Psi^{(2)} | \bigg( ({ g}_{0}^{(1)})^a{}_b | {\cal W}_1^{2,3} \rangle^{bc}_{ABC}
 -
 ({g}_{0}^{(2)})^c{}_b | {\cal W}_2^{1,3} \rangle^{ba}_{CBA}
+ 
{\tilde  Q}_{B}^{(3)} 
|{\cal V} \rangle^{ac}_{CAB}\bigg) =0
\ee

\be \label{xy-2}
{}^C\langle \phi^{(3)}|{}^A_a \langle \Psi^{(1)} | {}^B_c\langle \Lambda_F^{(2)} |\bigg( ({g}_{0}^{(1)})^a{}_b | {\cal W}_1^{3,2} \rangle^{bc}_{CBA}  
+ 
{ l}_{0}^{(3)}  
|{\cal W}_{3}^{1,2} \rangle^{ac}_{ABC}+
{\tilde Q}_{F}^{(2)} | {\cal V} \rangle^{ac}_{ABC}\bigg)
=0
\ee

\be \label{xy-3}
{}^C\langle \phi^{(3)}|{}^B_c \langle \Lambda_F^{(1)} | {}^A_a\langle \Psi^{(2)} |\bigg( ({g}_{0}^{(2)})^a{}_b | {\cal W}_2^{3,1} \rangle^{bc}_{CBA}
+ 
{ l}_{0}^{(3)}  
|{\cal W}_{3}^{2,1} \rangle^{ac}_{ABC}-
{\tilde Q}_{F}^{(1)} | {\cal V} \rangle^{ca}_{BAC}\bigg)
=0
\ee
where \(|\Psi^{(i)}\rangle\), \(|\phi^{(3)}\rangle\), \(|\Lambda_B^{(i)}\rangle\), and \(|\Lambda_F^{(3)}\rangle\) are constrained as described in section 2.

Furthermore, for the preservation of the group structure of the gauge transformations
up to the first order in the coupling constant $g$
there must exist some functions $| {\cal X}_i \rangle$ such that
\be \label{inv-w-1}
{}_b^A\langle \Lambda_F^{(2)} | {}^B\langle \Lambda_B^{(3)} |\bigg( {\tilde Q}_{F}^{(2)} | {\cal W}_1^{2,3} \rangle^{ab}_{ABC} 
+ \tilde Q_{B}^{(3)} | 
{\cal W}_{1}^{3,2} \rangle^{ab}_{BAC} -
{\tilde Q}_F^{(1)} | {\cal X}_1 \rangle^{ab}_{ABC}\bigg) = 0
\ee
\be \label{inv-w-2}
{}^B\langle \Lambda_B^{(3)} | {}^A_b\langle \Lambda_F^{(1)} |\bigg({
  \tilde Q}_{F}^{(1)} | {\cal W}_2^{1,3} \rangle^{ab}_{ABC} 
+ \tilde Q_{B}^{(3)} | 
{\cal W}_{2}^{3,1} \rangle^{ab}_{BAC} -
{\tilde Q}_F^{(2)} | {\cal X}_2 \rangle^{ab}_{ABC}\bigg) = 0
\ee
\be \label{inv-w-3}
 {}^A_a\langle \Lambda_F^{(1)} | {}^B_b\langle \Lambda_F^{(2)} |\bigg({\tilde Q}_{F}^{(1)}
| {\cal W}_3^{1,2} \rangle^{ab}_{ABC}
-
{\tilde Q}_{F}^{(2)}  
|{\cal W}_3^{2,1} \rangle^{ba}_{BAC}
-
\tilde Q_{B}^{(3)} 
|{\cal X}_{3} \rangle^{ab}_{ABC}\bigg) = 0
\ee
Since both the Lagrangian and the higher spin functionals have ghost number zero, it follows that the vertex $| {\cal V} \rangle$  has  the ghost number $0$ and 
the $| {\cal W} \rangle$ and $| {\cal X} \rangle$-- vertices have
ghost number $+1$. 

Let us note, that the Lagrangian \p{int.} is symmetric under the exchange of 
$|\Psi^{(1)}\rangle$ and $|\Psi^{(2)}\rangle$, provided the vertex obeys the symmetry  
\be 
1 \leftrightarrow 2, \qquad |{\cal V}\rangle^{ab}_{ABC} \to -|{\cal V}\rangle^{ba}_{BAC} \label{exchangesym} 
\ee
The   gauge transformation rules   \p{BFFG-1}-\p{BFFG-2-2} 
are symmetric under the exchange of labels $1$ and $2$ as well. Similarly, the transformation
 \p{exchangesym} leaves the equation \p{xy-1} invariant, and takes the equation \p{xy-2} to \p{xy-3} and vice versa.

To summarize, the vertex $| {\cal V} \rangle^{ab}$
describes the Lagrangian cubic interactions and the  $|{\cal W} \rangle^{ab}$--vertices describe  
nonlinear
deformations of the linear gauge transformations.
The defining equations are \p{xy-1}--\p{xy-3}, whereas the equations
\p{inv-w-1}--\p{inv-w-3}
are in a sense the consistency conditions for the vertices.

\section{Super Yang-Mills-like Systems} \label{ssym} \setcounter{equation}0

\subsection{Vertices}
Let us consider a vertex
\be \label{vertex-1}
({\cal V})^{ab}_{ABC} = f_{ABC} (\gamma \cdot \alpha_2^+ )^{ab} {\cal F} ({\cal K}^{(i)}_1, {\cal Z}_{111})
\ee
for cubic interactions between two fermions and one boson. The function ${\cal F}$ is an arbitrary function of ${\cal Z}_{111}$ and ${\cal K}_1^{(i)}$, as defined in \p{sbv-1-x} and \p{sbv-3}.We can solve explicitly equations \p{xy-1}-\p{inv-w-3} for any ${\cal F}$, with the solutions given in appendix \p{Appendix D}.

To simplify the following and aid in establishing supersymmetry we impose a cyclic symmetry on the vertex, by
choosing ${\cal F}$ such that
\be \label{cyclicF}
\frac{\partial {\cal F}}{\partial {\cal K}^{(1)}_1} = \frac{\partial {\cal F}}{\partial {\cal K}^{(2)}_1} = \frac{\partial {\cal F}}{\partial {\cal K}^{(3)}_1} \equiv \frac{\partial {\cal F}}{\partial {\cal K}_1} \ee

In order to consider $N=1$ supersymmetry we shall choose the following cubic vertex
for three bosonic higher spin fields
\be \label{vertex-1-3b}
   {}^C\langle \phi^{(1)} | \,\,
{}^A\langle \phi^{(2)}| \,\,
{}^B\langle \phi^{(3)}|  
{\cal Z}_{222}  {\cal F} ({\cal K}_1, {\cal Z}_{111})
| 0_B^{(1)} \rangle \otimes
|0_B^{(2)} \rangle \otimes | 0_B^{(3)} \rangle
f_{ABC} 
\ee
The interaction between two fermions and one boson
 is described by the cubic vertex \p{vertex-1}. However,
for the purpose of finding supersymmetry transformations, we take
three Fock spaces in the fermionic sector as well 
and consider the interactions between two fermions and one boson as 
\be \label{int-s-1}
  \,\,    {}^C\langle \phi^{(3)} | \,\,
{}_a^A\langle \Psi^{(1)}| \,\,
{}_b^B\langle \Psi^{(2)}| (\gamma \cdot \alpha_2^{(3),+} )^{ab}{\cal F}({\cal K}_1,{\cal Z}_{111})
| 0_F^{(1)} \rangle \otimes
|0_F^{(2)} \rangle \otimes | 0_B^{(3)} \rangle
f_{ABC}
+ \,\, \text{cyclic}
\ee
Given the symmetry of exchanging the Fock space labels \p{exchangesym} in the definition of the vertices,
the function \({\cal F}\) has to be even, in the sense that
\be {\cal F}(-{\cal K}_1,{-\cal Z}_{111}) = {\cal F}({\cal K}_1,{\cal Z}_{111}) \ee
This implies that the total number of \(\alpha_1^+\) oscillators in \({\cal F}\) is even, and the total number of oscillators in the vertices is odd.

Naturally, in order to establish supersymmetry for the nonlinear systems
under consideration,
one starts with the transformations that connect the free Lagrangians
for fermionic and bosonic (generalised) triplets \cite{Sorokin:2018djm}
\bea
&&\delta | \phi^{(i)} \rangle_A = \bar \epsilon_a \, (\alpha_{2}^{(i),+} \cdot \gamma)^a{}_b  
| \Psi^{(i)} \rangle^b_A ,
\label{susy-1-1} \\
&&\delta | \Psi^{(i)} \rangle^a_A  =  - 2 (p^{(i)} \cdot \gamma)^a{}_b \, 
(\alpha_{2}^{(i)}  \cdot \gamma)^b{}_c \, \epsilon^c \,
| \phi^{(i)} \rangle_A   \label{susy-2-1}
\eea
and then considers  their nonlinear deformations by the terms which are compatible
with the interactions.\footnote{From these supersymmetry transformations one can see
that  the fields $\phi_{\nu; \mu_1,...,\mu_n}(x)$ and $\Psi_{_{ \mu_1,...,\mu_n}}(x)$
form an $N=1$ supermultiplet, see \cite{Sorokin:2018djm} for details.}
One can see, however, that the off-shell transversality conditions
\p{GFFF} and \p{GENTR} combined with supersymmetry transformations
\p{susy-1-1}--\p{susy-2-1} puts the fields completely on shell.
On the other hand, it is a matter of direct computations to check that the supersymmetry transformations
given above transform the vertex \p{vertex-1-3b}
into  \p{int-s-1}  and vice versa, provided the fields are transversal
and obey the massless Klein-Gordon and Dirac equations.
This invariance can be explained as follows:
in the case of free triplets \cite{Sorokin:2018djm}
supersymmetry transformations are generated by the oscillator $\alpha_{2}^{(i),+}$, see 
\p{susy-1-1}--\p{susy-2-1}.
Since the fields are on shell, these transformations stay the same
also for cubic interactions.
Further,  both vertices \p{vertex-1-3b}
and
\p{int-s-1}
have  the form of an unknown function 
which depends 
 only on the 
oscillators $\alpha_{1}^{(i),+}$, times  prefactors which contain
only  the oscillators $\alpha_{2}^{(i),+}$.
Therefore, it is sufficient  to check how these prefactors transform into each other
under the supersymmetry transformations. As one can see,  this check repeats exactly the  proof
for the invariance  of cubic interactions  in the standard  $N=1$ Super Yang-Mills theory.

\subsection{An Example: \texorpdfstring{$N=1$}{N=1} Super Yang-Mills Theory} \label{sec:SYM}
Cubic vertices  for  $N=1$ super Yang-Mills theory
in $D=4,6$ and $10$ dimensions are the simplest examples of the 
ones considered in the previous subsection.
We shall consider them in detail 
also for the purpose of showing how
the requirement imposed by supersymmetry for the fields being completely on shell can be lifted 
by including auxiliary fields, thus promoting the system to an off-shell one.

 For the case of super Yang-Mills theory we take at most only one set of oscillators
$\alpha^{\mu,+}_2, c_2^+, b_2^+$ in each Fock space, thus
making the nonlinear deformation of the Super-Maxwell system considered in \cite{Sorokin:2018djm}.

To obtain the Yang-Mills cubic vertex we take 
the higher spin functional in the form
\be \label{hsf-b-ym}
| \phi^{(i)} \rangle_A = {\cal A}_{\mu,A}(x) \alpha_2^{\mu (i), +}|0^{ (i)}_B \rangle
\ee
and then use \p{vertex-1-3b} with the unknown function ${\cal F}$ being replaced by a constant
\bea \label{YM-111}
| V \rangle_{ABC}&=& -\frac{ig}{12} f_{ABC} 
 {\cal Z}_{222} | 0^{(1)}_B \rangle \otimes
|0^{(2)}_B\rangle \otimes | 0^{(3)}_B \rangle
\eea
In this way one obtains the cubic interaction vertex of Yang-Mills theory
\be
V= g f_{ABC}(\partial^\mu {\cal A}^{\nu}_{A}) {\cal A}_{\mu,B} {\cal A}_{\nu,C}
\ee

Similarly, we take the higher spin functional in the fermionic sector as
\be \label{psi-YM-1}
| \Psi^{(i)} \rangle^a_A =  
\Psi^{a}_A (x) |0_F^{(i)}\rangle
\ee
Then, from the vertex \p{int-s-1}, with constant ${\cal F}$ 
\be \label{YM-222}
|{\cal V} \rangle^{ab}_{ABC} = \frac{i g}{3} f_{ABC}(\alpha_2^{(3),+} \cdot \gamma)^{ab} \,\, 
| 0^{(1)} \rangle_{F} \otimes
|0^{(2)} \rangle_{F} \otimes | 0^{(3)} \rangle_{B} + \,\, \text{cyclic}
\ee
we get for the cubic interaction between two fermions and the gauge field
\be
{\cal V} = i g f_{ABC} \Psi^{a,A} \gamma^\mu_{ab} \Psi^{b,B} {\cal A}^C_\mu 
\ee
The only nonzero parameter of gauge transformations is
\be
| \Lambda_{B}^{(i)} \rangle^A = i b_2^{(i),+} \lambda^A(x) | 0_B^{(i)} \rangle.
\ee
From the equations \p{s-w-ym-6-1} and \p{s-w-ym-6} with constant ${\cal F}$
we get for the nonzero components of ${\cal W}$ vertices
\be
| {\cal W}_1^{2,3} \rangle^{ab}_{ABC} =
| {\cal W}_2^{1,3} \rangle^{ab}_{ABC}
=f_{ABC}c_2^+ C^{ab}
 | 0^{(1)}_F \rangle \otimes
|0^{(2)}_F \rangle \otimes | 0^{(3)}_B \rangle,
\ee
which generate the standard gauge transformations for spin $1/2$ fermions in the adjoint representation
\be
\delta \Psi^{a}_A = g f_{ABC} \Psi^{a}_B \lambda_C
\ee
Then using \p{susy-1-1}--\p{susy-2-1}
we get the linear part of the standard supersymmetry transformations for the Yang-Mills 
supermultiplet
\be
\delta {\cal A}_{\mu, A} = i \Psi^{a}_A(\gamma_\mu)_{ab} \epsilon^b,
\quad
\delta {\Psi}^{a}_A = i  (\gamma^\nu)^a{}_b (\gamma^\mu)^b{}_c \,
\epsilon^c \partial_\nu {\cal A}_{\mu, A}. 
\ee
The equation above describes on-shell vertices.
To promote this system off-shell, 
instead of imposing an off-shell transversality constraint,
we introduce an auxiliary ${\cal E}^A(x)$ field.
That means, that in the bosonic sector we consider a higher spin functional
 of the form
\be \label{YM-333}
| \Phi^{(i)}_{B} \rangle^A = ({\cal A}_\mu^A(x) \alpha_2^{\mu (i), +}
-i {\cal E}^A(x) c_0^{(i), + } b_2^{(i), + })|0^{(i)}_B  \rangle.
\ee
 The expression for the
interaction vertex between two fermions and the boson  remains unchanged, 
while the interaction vertex between three bosons we now write as
\bea \label{YM-vvv} \nonumber
| V \rangle_{ABC}&= -\frac{i g}{12} f_{ABC} [(\alpha_2^{(1),+} \cdot \alpha_2^{(2),+})((p^{(1)}- p^{(2)}) \cdot \alpha_2^{(3),+} +
(b_0^{(1)}-b_0^{(2)}) \, c_2^{(3),+}) ] \times \\ 
&\qquad\times c_0^{(1)}c_0^{(2)}c_0^{(3)}
\,\, 
| 0^{(1)}_B \rangle \otimes
|0^{(2)}_B \rangle \otimes | 0^{(3)}_B \rangle + \,\, \text{cyclic}
\eea
The full interacting cubic Lagrangian  is a sum of
\p{L-bbb-gf}
and of
\be\label{LNSINT-YM-2}
{ \cal L}_{\text{int}} =\sum_{i=1}^3 
{}^A\langle \Psi^{(i)} |g_{0}^{(i)}| \Psi^{(i)}
\rangle_A + 
g \left ( {}^A\langle \Psi_{F}^{(1)}| {}^B\langle \Psi^{(2)}| {}^C\langle \Phi_{B}^{(3)}| 
|{\cal V} \rangle_{ABC}  + \,\, \text{cyclic}  \right )
\ee
Using the explicit form 
of the higher spin functionals in the bosonic \p{YM-333} and fermionic \p{psi-YM-1} sectors,
one can see that the auxiliary field 
$ {\cal E}^A(x)$  is contained only in the free part of the bosonic Lagrangian.
After eliminating it via its own equations of motion
one obtains the standard Lagrangian for super Yang-Mills up to the  cubic order.

Let us note that the vertex \p{YM-vvv} generates  both the Lagrangian interactions and the nonlinear part of the gauge transformations
\be
\delta {\cal A}_{\mu, A} = \partial_\mu \lambda_A + g f_{ABC} {\cal A}_{\mu,B} \lambda_C.
\ee
The supersymmetry transformations 
will be deformed with nonlinear terms
\bea \label{SUSY-sym}
&&\delta | \phi^{(i)} \rangle_A = \bar \epsilon_a \, (\alpha_{2}^{(i),+} \cdot \gamma)^a{}_b  
| \Psi^{(i)} \rangle^{b}_A, \label{susy-1-n} \\
&&\delta | \Psi^{(i)} \rangle^{a}_A  =  - 2 (p^{(i)} \cdot \gamma)^a{}_b \, (\gamma^\mu)^{b}{}_c
\, \alpha_{2,\mu}^{(i)} \, \epsilon^c \,
| \phi^{(i)} \rangle_A +  \label{susy-2-n}
\\ \nonumber
&&+g \, 
{}_B\langle \phi^{(i+1)}| {}_C\langle \phi^{(i+2)}| f_{ABC}
(\gamma^{\mu \nu})^a{}_b\alpha^{(i+1),+}_{2,\mu}  \alpha^{(i+2),+}_{2,\nu} \epsilon^b \,\,
| 0^{(i+1)}_B \rangle \otimes
|0^{(i+2)}_B \rangle \otimes | 0^{(i)}_F \rangle
\eea
being the standard supersymmetry transformations for the $N=1$  Yang-Mills supermultiplet.

The consideration of the Super Yang-Mills theory suggests a very interesting possibility
to lift the supersymmetry to an off-shell Lagrangian level,
 by considering the gauge fixing condition \p{GFFF} only for $m=1$. In other words, after imposing transversality only
with respect to the first set of the indices, the supersymmetry no longer requires
the higher spin fields to be on-shell, and since they contain only one 
$\alpha_2^{(i),+}$ oscillator (like  the Yang-Mills vector field), the nonlinear part of the supersymmetry transformations will be the same
as in \p{SUSY-sym}. Exactly the same arguments can be applied to the supergravity-like systems
considered below, with  nonlinear parts of supersymmetry transformations being determined by the
corresponding $N=1$ supergravity transformations.

\section{Supergravity-like systems} \label{ssugra} \setcounter{equation}0
\subsection{Vertices}
As the second type of the vertices we consider the case where the internal indices are absent. The systems obtained in this way lead to the higher spin generalization of $D=4$
$N=1$ supergravity, as we shall see below.

The defining equations and consistency conditions 
for the supegravity-like vertices
are again \p{xy-1}--\p{xy-3} and \p{inv-w-1}--\p{inv-w-3}.
One can see that the corresponding solutions
for ${\cal W}$ vertices can be obtained from the ones for the vertex \p{vertex-1}
by simply flipping the signs
of $({\cal W}_1^{3,2})^{ab}$, $({\cal W}_2^{1,3})^{ab}$ and $({\cal W}_3^{2,1})^{ab}$, since now we do not have to account for the antisymmetry of the structure constants.

We consider the following interaction vertex between two fermionic and one bosonic fields 
\be \label{sugralikebff}
  \,\,    \langle \phi^{(3)} | \,\,
{}_a\langle \Psi^{(1)}| \,\,
{}_b\langle \Psi^{(2)}| {\cal Z}_{111}(\gamma \cdot \alpha_2^{(3),+} )^{ab}{\cal F}({\cal K}^{(i)}_1,{\cal Z}_{111})
| 0_F^{(1)} \rangle \otimes
|0_F^{(2)} \rangle \otimes | 0_B^{(3)} \rangle
+ \,\, \text{cyclic}
\ee
The cubic vertex for three bosonic fields is
\be \label{sugralike3b}
   \langle \phi^{(1)} | \,\,
\langle \phi^{(2)}| \,\,
\langle \phi^{(3)}| {\cal Z}_{111}
{\cal Z}_{222}  {\cal F} ({\cal K}^{(i)}_1, {\cal Z}_{111})
| 0_B^{(1)} \rangle \otimes
|0_B^{(2)} \rangle \otimes | 0_B^{(3)} \rangle.
\ee
Apart from the absence of internal indices
the difference from the super Yang-Mills-like vertex is  the inclusion of ${\cal Z}_{111}$ in the prefactor. As before, the undetermined arbitrary function in the vertices can depend on ${\cal K}^{(i)}_1$ and ${\cal Z}_{111}$, but we impose cyclicity using condition \p{cyclicF}.

The consideration of $N=1$ supersymmetry closely resembles the one for the Super Yang-Mills -like systems. The supersymmetry transformations are \p{susy-1-1}--\p{susy-2-1} without internal symmetry indices. They take the vertices \p{sugralikebff} and \p{sugralike3b} to each other, with the proof being completely analogous to the one used 
in Super Yang-Mills. Again the transversality constraint puts the fields completely on-shell.

The generalizing function has to be even
\be {\cal F}(-{\cal K}_1, -{\cal Z}_{111}) = {\cal F}({\cal K}_1, {\cal Z}_{111}) \ee
otherwise  the vertex evaluates to zero because of the symmetry of changing the Fock space labels. This means that the total number of oscillators in the supergravity-like vertices is even.

\subsection{An Example: \texorpdfstring{$D=4$, $N=1$}{D=4 N=1} supergravity}\label{gr+b}

In this section we shall demonstrate how  the present approach works for the case of
the linearized $D=4$, $N=1$ supergravity.  

Following \cite{Sorokin:2018djm}, let us take the higher spin functional to contain two oscillators
in the bosonic sector
\be
| \phi^{(i)} \rangle =  \phi_{\mu, \nu}(x) \,\alpha_{1}^{\mu (i),+} \alpha_2^{\nu (i),+} 
|0_B^{ (i)} \rangle
\ee
with no symmetry between the two indices, and one oscillator in the fermionic sector
\be
| \Psi^{(i)}\rangle^a = \Psi_\mu^{a}(x) \,  \alpha_1^{\mu (i),+} | 0_F^{ (i)} \rangle
\ee
Decomposing the fields into irreducible representations of the Poincar\'e group as
\be \label{dec-1}
\phi_{\mu, \nu}  = \left ( \phi_{(\mu, \nu )} - \eta_{\mu \nu}\frac{1}{D} \phi_\rho^\rho
\right ) + \phi_{[\mu, \nu]} +
 \eta_{\mu \nu}\frac{1}{D}  \phi_{\rho}^\rho
\equiv
h_{\mu \nu} + B_{\mu \nu} + \frac{1}{D}\eta_{\mu \nu} \varphi
\ee
and
\be \label{dec-2}
\psi_\mu^a = \Psi_\mu^a +\frac{1}{D}(\gamma^\mu)^{ab} (\gamma^\nu)_{bc} \Psi_\nu^c
\equiv \Psi_\mu^a + \frac{1}{D}(\gamma^\mu)^{ab} \Xi_b
\ee
one can see, that this field content corresponds to the $D=4$ $N=1$ supergravity supermultiplet and a chiral
supermultiplet \cite{Cremmer:1978hn}.

The interaction vertices are given by \p{sugralikebff} and \p{sugralike3b} with the function ${\cal F}$ 
 being a constant. Taking the cubic interaction vertex between three bosons as
\be \label{sugra3b}
V = \frac{1}{6} a {\cal Z}_{111} {\cal Z}_{222},
\ee
where the expressions for ${\cal Z}_{mnp}$ are given in
\p{sbv-3}, one obtains the Lagrangian
\be \label{t2}
{ \cal L}_{B}=  - \phi^{\mu, \nu} \Box \phi_{\mu,\nu }
-4a ( \partial_\rho \partial_\tau \phi_{\mu, \nu}) \phi^{\mu, \nu} \phi^{\rho, \tau }
+ 8 a (\partial_\rho \partial_\tau \phi_{\mu, \nu}) \phi^{\mu, \tau} \phi^{\rho, \nu}
\ee
The vertex describing interactions between two fermionic and one bosonic fields
is
\be \label{ffb-sugra-2}
\frac{f}{3}\left(\langle \phi^{(3)}| {}_a\langle \Psi^{(1)}| {}_b \langle \Psi^{(2)}| (\gamma_\mu)^{ab} \alpha_2^{\mu (3), +} {\cal Z}_{111}|0_F^{(1)}\rangle \otimes |0_F^{(2)}\rangle \otimes |0_B^{(3)}\rangle + \text{cyclic}\right)
\ee
From these vertices we obtain the Lagrangian for the fermions
\be \label{t3}
{ \cal L}_F = -\frac12 {\bar\Psi}^\mu \gamma^\nu \partial_\nu \Psi_\mu + 4if \phi^{\mu,\nu} {\bar \Psi}^\alpha \gamma_\nu \partial_\alpha \Psi_\mu - 2if \phi^{\mu,\nu} {\bar \Psi}^\alpha \gamma_\nu \partial_\mu \Psi_\alpha
\ee
One can expand \p{t2} and \p{t3} and write the Lagrangian in terms of the irreducible components.

In order to consider $N=1$ supersymmetry,
we restrict the fields to be completely on-shell, 
as we have done for the Yang-Mills like systems.
A choice of the constants as $a = -4f$ allows one to match the relative coefficients between the cubic vertices to the one of the linearized pure $D=4$ $N=1$ supergravity
(see  Appendix \ref{AppendixB} for some equations for  linearized supergravity).
Then one can check that the transformations \p{susy-1-1}--\p{susy-2-1}
transform the cubic vertices  \p{sugra3b} and \p{ffb-sugra-2}
 into each other.
The supersymmetry transformations for irreducible components
can be read from \p{susy-1-1}--\p{susy-2-1} and
correspond to  supersymmetry trasnformations
of  the linearized $D=4$ $N=1$ Supergravity  \cite{Sorokin:2018djm}.

Let us note that the field content \p{dec-1}--\p{dec-2} corresponds also to the irreducible
$N=1$ supergravity supermultiplet in ten dimensions \cite{Chamseddine:1980cp}--\cite{Bergshoeff:1981um}
and to $N=(1,0)$ gravitational supermultiplet
together with
$N=(1,0)$ tensor supermultiplets\footnote{The six-dimensional $N=(1,0)$ gravitational supermultiplet
$(h_{\mu \nu}, B_{\mu \nu}^+, \psi_\mu)$ contains a graviton, the self-dual part
of the $B_{\mu \nu}$ field and a chiral gravitino. 
The six-dimensional $N=(1,0)$ tensor supermultiplet 
$(\phi, B_{\mu \nu}^-, \Xi)$  contains a scalar, the anti-self-dual part of the $B_{\mu \nu}$ field and an 
anti-chiral fermion.}
in six dimensions
\cite{DallAgata:1997yxl}.
However, although the invariance under supersymmetry transformations
works exactly in the same way as for four dimensions,
a promotion of these higher dimensional models 
to off-shell ones  for higher spin fields might prove problematic. The reason 
for this is that it does not seem possible 
to find a consistent higher spin generalization of the vertices
which describe  the coupling of the $B_{\mu \nu}$ field
to the fermions in the $D=10$, $N=1$ supergravity \cite{Chamseddine:1980cp}--\cite{Bergshoeff:1981um} given in equation \p{vertex_10dSugraB}.
Therefore, an off-shell higher spin extension of higher dimensional supergravity-like 
models still poses an interesting open problem.

\section{Conclusions} \setcounter{equation}0 \label{Conclusions}
In this paper we have constructed the cubic interaction vertices for the massless  higher spin supersymmetric theories in four, six and ten dimensions. Our analysis is based on use of the BRST approach to higher spin field  theories which works perfectly both for finding the free Lagrangians and the vertices. As a concrete application we have studied the vertices for Yang-Mills-like higher spin theories which are characterised by Lie algebra structure, and for $N=1$ supegravities in $D=4$, $6$ and $10$. 

The present paper is a step towards
off-shell Lagrangian formulation
of supersymmetric higher spin gauge theories in various dimensions.
Since  computations for the  off-shell
unconstrained Lagrangians are quite tedious, here we restricted ourselves
with the consideration of maximally simplified models. In particular,
we started with unconstrained free Lagrangians for massless reducible representations
of the Poincar\'e group 
and  gauge fixed
them to contain only d'Alembertian and Dirac operators, while 
 keeping the transversality conditions off-shell.
As a second step, we considered the cubic interactions for such Lagrangians.
Finally, we showed that supersymmetry transformations, under which the obtained system
is invariant, put the fields completely on-shell. 
All these steps, however, can be  generalised to an 
unconstrained off-shell form via straightforward computations, which 
we leave to a separate publication. It is interesting to note that in four dimensions the most convenient way to develop the unconstrained formulation is one in terms of two-component totally symmetric spin tensors where the trace conditions are automatically fulfilled (see e.g. \cite{Buchbinder:2015kca}--\cite{Buchbinder:2020yip}).

It would be interesting to consider massive higher spin supermultiplets in higher dimensions, a topic which to the best of our knowledge has not been yet explored. Further inclusion of cubic interactions into these systems is not only interesting in its own right, but hopefully might shed some new light on the role played by massive higher spin modes in superstring theories (see \cite{Marotta:2021oiw} for a recent study in this direction).

A possible deformation of the models presented in the present paper to curved backgrounds is yet  another interesting problem. For this purpose  $AdS_D$ space is a natural
choice, since it is generically compatible with supersymmetry, unlike de Sitter spaces (see \cite{David:2020ptn} for a recent discussion on higher spin theories on $dS_4$).
Again, despite the recent progress in studies of supersymmetric higher spin models on $AdS$ backgrounds \cite{Hutomo:2019mcx}--\cite{Hutchings:2020eff}, the higher dimensional generalizations are not known.
 
Most importantly, it is interesting to find if there is a possibility for building supersymmetric models
which have consistent higher order classical, and possibly quantum, interactions. 
It is well known that real difficulties in higher spin theories
 start when considering higher order interactions, even at the classical level\footnote{For 
the non-supersymmetric case this problem can be overcome by considering 
 four or three dimensional theories in the light-front gauge 
 with the coupling constants in the cubic vertices having a specific form 
 \cite{Metsaev:1991mt}--\cite{Skvortsov:2020pnk}. A detailed study of the quantum properties of these models have been performed in \cite{Skvortsov:2018jea}--\cite{Skvortsov:2020gpn}.}. It would be very interesting, therefore, to explore the possibility of the existence of supersymmetric theories with massive and/or massless higher spin fields with consistent higher order classical and quantum interactions.

\vskip 0.5cm

\noindent {\bf Acknowledgments.} We are
grateful to Yasha Neiman and Dmitri Sorokin
for useful discussions.
The work of I.L.B. and V.A.K. was partially supported by The Ministry of Education of Russian Federation, project FEWF-2020-0003.  The work of M.T. and D.W. was supported by the Quantum Gravity Unit
of the Okinawa Institute of Science and Technology Graduate University
(OIST).

\renewcommand{\thesection}{A}

\renewcommand{\theequation}{A.\arabic{equation}}

\setcounter{equation}0
\appendix
\numberwithin{equation}{section}

\section{Conventions}\label{Appendix A}
We mainly  follow the notations of \cite{VanProeyen:1999ni}, where some more useful identities
for spinors and for gamma matrices can be found.

Throughout the paper  ``$(,)$'' denotes symmetrization and ``$[,]$'' denotes antisymmetrization with weight one.
The Latin letters $a,b \ldots$ label spinorial indices.  The Greek letters
$\mu, \nu, \ldots$ label flat space-time vector indices
and  Greek letters with ``hat'' ${\hat \mu}, {\hat \nu},\ldots$ label vector indices in curved space-time.

We choose a real representation for Majorana spinors
\be
(\lambda^a)^\star=\lambda^a, \quad \bar \lambda_a =\lambda^b C_{ba}
\ee
The spinor indices can be raised and lowered by anti-symmetric charge conjugation matrices $C_{ab}$ and $C^{ab}$  as
\be
\lambda^a = C^{ab} \lambda_b, \quad \lambda_a = \lambda^b C_{ b a},
\quad C^{ab}C_{bc}=-\delta^a_c.
\ee
The $\gamma$--matrices satisfy the following anti-commutation relations
\begin{equation}\label{1}
(\gamma^\mu)^a{}_c (\gamma^\nu)^c{}_b
+
(\gamma^\nu)^a{}_c (\gamma^\mu)^c{}_b
 = 2 \eta^{\mu \nu} \delta^a_b.
\end{equation}
 In $D=4$ the matrices $\gamma_\mu$ and $\gamma_{\mu \nu}$ with both spinorial indices  up (down)
 are symmetric and the matrices $C$, $\gamma_5$ and $\gamma_5 \gamma_\mu$ are anisymmetric. In $D=10$ the matrices
 $\gamma_\mu$ and $\gamma_{\mu_1,...,\mu_5}$ with both spinorial indices up (down)
 are symmetric, and the matrices $\gamma_{\mu_1 \mu_2 \mu_3}$ are antisymmetric.

For checking the on-shell closure of the supersymmetry algebra and of the 
supersymmetry of the vertices we have used the following gamma-matrix identities
\be\label{FI}
(\gamma^\nu)_{ab}{(\gamma_ \nu)}_{ cd}+ (\gamma^\nu)_{ac}{(\gamma_ \nu)}_{ db} + (\gamma^\nu)_{ad}{(\gamma_ \nu)}_{ bc}=0,
\ee
\be
\gamma^\mu \gamma^{\nu_1, \nu_2, ... \nu_r} \gamma_\mu=
(-1)^r (D-2r) \gamma^{\nu_1, \nu_2,...\nu_r}.
\ee
For a product of gamma matrices we have
\be
\gamma^{\nu_1,..., \nu_i} \gamma_{\mu_1,..., \mu_j}=
\sum_{k =0}^{k =min (i,j)} \frac{i! j!}{(i-k)! (j-k)! k!} 
\gamma^{[\nu_1,...,\nu_{i-k}}{}_{[\mu_{k+1},...,\mu_{j}} \delta^{\nu_i}_{\mu_1} \delta^{\nu_{i-1}}_{\mu_2}...
\delta^{\nu_{n-k+1}]}_{\mu_k]}
\ee
and in particular
\be \label{GG-1}
\gamma_{\mu \nu \rho}= \gamma_{\mu \nu} \gamma_\rho -2 \eta_{\rho [\nu} \gamma_{\mu]}.
\ee

\section{Equations for \texorpdfstring{${\cal W}$}{W} vertices}\label{Appendix C}
The equations which express the ${\cal W}$ vertices in terms
of the ${\cal V}$ vertices present
in the Lagrangian \p{L-ffb-gen} are\bea \nonumber \label{eq-unf-1}
 \int dc_0^{(3)} &\bigg(& Q^{(3)}| {\cal V}_{11}\rangle^{ab}_{ABC} 
-\frac{1}{\sqrt 2} (g_0^{(1)})^a{}_c |{\cal W}_{1,11}^{2,3}\rangle^{cb}_{BCA}
+\frac{1}{\sqrt 2}(g_0^{(2)})^b{}_c |{\cal W}_{2,11}^{1,3}\rangle^{ca}_{ACB} + \\ 
&-& \tilde Q^{(1)} |{\cal W}_{1,21}^{2,3}\rangle^{ab}_{BCA}
+  \tilde Q^{(2)} |{\cal W}_{2,21}^{1,3}\rangle^{ba}_{ACB}
\bigg) = 0\eea

\bea \nonumber
 \int dc_0^{(3)} &\bigg(& Q^{(3)}| {\cal V}_{22}\rangle^{ab}_{ABC} 
+{\sqrt 2} M_F^{(1)}(g_0^{(1)})^a{}_c |{\cal W}_{1,22}^{2,3}\rangle^{cb}_{ACB}
+{\sqrt 2} M_F^{(2)}(g_0^{(2)})^b{}_c |{\cal W}_{2,22}^{1,3}\rangle^{ca}_{BCA} + \\ 
&+& \tilde Q^{(1)} |{\cal W}_{1,12}^{2,3}\rangle^{ab}_{BCA}
+  \tilde Q^{(2)} |{\cal W}_{2,12}^{1,3}\rangle^{ba}_{ACB}
\bigg) = 0\eea

\bea \nonumber \label{eq-unf-3}
 \int dc_0^{(3)} &\bigg(& Q^{(3)}| {\cal V}_{12}\rangle^{ab}_{ABC} 
+\frac{1}{\sqrt 2} (g_0^{(1)})^a{}_c |{\cal W}_{1,12}^{2,3}\rangle^{cb}_{BCA}
-{\sqrt 2} M_F^{(2)}(g_0^{(2)})^b{}_c |{\cal W}_{2,21}^{1,3}\rangle^{ca}_{ACB} + \\ 
&-& \tilde Q^{(1)} |{\cal W}_{1,22}^{2,3}\rangle^{ab}_{BCA}
+  \tilde Q^{(2)} |{\cal W}_{2,11}^{1,3}\rangle^{ba}_{ACB}
\bigg) = 0\eea

\bea \nonumber \label{eq-unf-4}
 \int dc_0^{(3)} &\bigg(& Q^{(3)}| {\cal V}_{21}\rangle^{ab}_{ABC} 
+\frac{1}{\sqrt 2} (g_0^{(2)})^b{}_c |{\cal W}_{2,12}^{1,3}\rangle^{ca}_{ACB}
-{\sqrt 2} M_F^{(1)}(g_0^{(1)})^a{}_c |{\cal W}_{1,21}^{2,3}\rangle^{cb}_{BCA} + \\ 
&+& \tilde Q^{(1)} |{\cal W}_{1,11}^{2,3}\rangle^{ab}_{BCA}
-\tilde Q^{(2)} |{\cal W}_{2,22}^{1,3}\rangle^{ba}_{ACB}
\bigg) = 0\eea

\bea \nonumber \label{eq-unf-5}
 \int dc_0^{(3)}&\bigg(&
 \tilde Q_F^{(1)} | {\cal V}_{11}\rangle^{ab}_{ABC} 
 + \frac{1}{\sqrt 2}(g_0^{(1)})^a{}_c |{\cal V}_{21}\rangle^{cb}_{ABC}
 + Q_B^{(3)} |{\cal W}_{3,11}^{2,1}\rangle^{ba}_{BAC} + \\
 &+& \frac{1}{\sqrt 2} (g_0^{(2)})^b{}_c |{\cal W}_{2,11}^{3,1}\rangle^{ca}_{CAB}
 - \tilde Q_F^{(2)} |{\cal W}_{2,21}^{3,1}\rangle^{ba}_{CAB}
\bigg) = 0\eea

\bea \nonumber
 \int dc_0^{(3)} &\bigg(&
 \tilde Q_F^{(1)} | {\cal V}_{12}\rangle^{ab}_{ABC} 
 - \frac{1}{\sqrt 2}(g_0^{(1)})^a{}_c |{\cal V}_{22}\rangle^{cb}_{ABC}
 + Q_B^{(3)} |{\cal W}_{3,21}^{2,1}\rangle^{ba}_{BAC} + \\
 &+& {\sqrt 2} M_F^{(2)}(g_0^{(2)})^b{}_c |{\cal W}_{2,21}^{3,1}\rangle^{ca}_{CAB}
 - \tilde Q_F^{(2)} |{\cal W}_{2,11}^{3,1}\rangle^{ba}_{CAB}
\bigg) = 0\eea

\bea \nonumber
 \int dc_0^{(3)} &\bigg(&
 \tilde Q_F^{(1)} | {\cal V}_{21}\rangle^{ab}_{ABC} 
 + {\sqrt 2}M_F^{(1)}(g_0^{(1)})^a{}_c |{\cal V}_{11}\rangle^{cb}_{ABC}
 + Q_B^{(3)} |{\cal W}_{3,12}^{2,1}\rangle^{ba}_{BAC} + \\
 &+& \frac{1}{\sqrt 2} (g_0^{(2)})^b{}_c |{\cal W}_{2,12}^{3,1}\rangle^{ca}_{CAB}
 + \tilde Q_F^{(2)} |{\cal W}_{2,22}^{3,1}\rangle^{ba}_{CAB}
\bigg) = 0\eea

\bea \nonumber \label{eq-unf-8}
 \int dc_0^{(3)} &\bigg(&
 \tilde Q_F^{(1)} | {\cal V}_{22}\rangle^{ab}_{ABC} 
 - {\sqrt 2}M_F^{(1)}(g_0^{(1)})^a{}_c |{\cal V}_{12}\rangle^{cb}_{ABC}
 + Q_B^{(3)} |{\cal W}_{3,22}^{2,1}\rangle^{ba}_{BAC} + \\
 &-& {\sqrt 2} M_F^{(2)}(g_0^{(2)})^b{}_c |{\cal W}_{2,22}^{3,1}\rangle^{ca}_{CAB}
 - \tilde Q_F^{(2)} |{\cal W}_{2,12}^{3,1}\rangle^{ba}_{CAB}
\bigg) = 0\eea

\bea \nonumber \label{eq-unf-9}
 \int dc_0^{(3)}&\bigg(&
 \tilde Q_F^{(2)} | {\cal V}_{11}\rangle^{ab}_{ABC} 
 - \frac{1}{\sqrt 2}(g_0^{(2)})^b{}_c |{\cal V}_{12}\rangle^{ac}_{ABC}
 - Q_B^{(3)} |{\cal W}_{3,11}^{1,2}\rangle^{ab}_{ABC} + \\
 &-& \frac{1}{\sqrt 2} (g_0^{(1)})^a{}_c |{\cal W}_{1,11}^{3,2}\rangle^{cb}_{CBA}
 + \tilde Q_F^{(1)} |{\cal W}_{1,21}^{3,2}\rangle^{ab}_{CBA}
\bigg) = 0\eea

\bea \nonumber
 \int dc_0^{(3)} &\bigg(&
 \tilde Q_F^{(2)} | {\cal V}_{21}\rangle^{ab}_{ABC} 
 - \frac{1}{\sqrt 2}(g_0^{(2)})^b{}_c |{\cal V}_{22}\rangle^{ac}_{ABC}
 + Q_B^{(3)} |{\cal W}_{3,21}^{1,2}\rangle^{ab}_{ABC} + \\
 &+& {\sqrt 2} M_F^{(1)}(g_0^{(1)})^a{}_c |{\cal W}_{1,21}^{3,2}\rangle^{cb}_{CBA}
 - \tilde Q_F^{(1)} |{\cal W}_{1,11}^{3,2}\rangle^{ab}_{CBA}
\bigg) = 0\eea

\bea \nonumber
 \int dc_0^{(3)} &\bigg(&
 \tilde Q_F^{(2)} | {\cal V}_{12}\rangle^{ab}_{ABC} 
 - {\sqrt 2}M_F^{(2)}(g_0^{(2)})^b{}_c |{\cal V}_{11}\rangle^{ac}_{ABC}
 + Q_B^{(3)} |{\cal W}_{3,12}^{1,2}\rangle^{ab}_{ABC} + \\
 &+& \frac{1}{\sqrt 2} (g_0^{(1)})^b{}_c |{\cal W}_{1,12}^{3,2}\rangle^{cb}_{CBA}
 + \tilde Q_F^{(1)} |{\cal W}_{1,22}^{3,2}\rangle^{ab}_{CBA}
\bigg) = 0\eea

\bea \nonumber \label{eq-unf-12}
 \int dc_0^{(3)} &\bigg(&
 \tilde Q_F^{(2)} | {\cal V}_{22}\rangle^{ab}_{ABC} 
 - {\sqrt 2}M_F^{(2)}(g_0^{(2)})^b{}_c |{\cal V}_{12}\rangle^{ac}_{ABC}
 + Q_B^{(3)} |{\cal W}_{3,22}^{1,2}\rangle^{ab}_{ABC} + \\
 &-& {\sqrt 2} M_F^{(1)}(g_0^{(1)})^a{}_c |{\cal W}_{1,22}^{3,2}\rangle^{cb}_{CBA}
 - \tilde Q_F^{(1)} |{\cal W}_{1,12}^{3,2}\rangle^{ab}_{CBA}
\bigg) = 0\eea

The ghost numbers for each vertex and of each  parameter of the gauge transformations
can be easily deduced,  using the following counting: the ghost number of the BRST charges
is equal to $+1$, integration over the ghost zero mode $c_0^{(3)}$ carries the ghost number
$-1$, the Lagrangian has the ghost number zero.
Then, the free part of \p{L-ffb-gen} implies that the fields
$|\Phi_{F,1}^{(i)}\rangle^b_A$ and $|\Phi_{B}^{(i)}\rangle_A$ have ghost number zero
and the field
$|\Phi_{F,2}^{(i)}\rangle^b_A$ has ghost number $-1$.
Similarly, the gauge transformation rules \p{BFFG-1-gen}--\p{BFFG-2-2-gen} and
the cubic part of the Lagrangian \p{L-ffb-gen}
determine the ghost numbers of the parameters of gauge transformations and 
of the vertices.

Of the above twelve equations, seven have a unique form. The remaining five are related to the others by utilizing the symmetry exchanging the Hilbert space labels for the fermions \(|\Phi_F^{(1)}\rangle\) and \(|\Phi_F^{(2)}\rangle\), together with the appropriate transformation of the vertex:
\be 1 \leftrightarrow 2\,,\qquad |{\cal V}_{mn}\rangle^{ab}_{ABC} \to (-1)^{mn}|{\cal V}_{nm}\rangle^{ba}_{BAC}. \ee
This operation will take the equation \p{eq-unf-3} to the equation \p{eq-unf-4} and, respectively, the equations \p{eq-unf-5}-\p{eq-unf-8} to the equations \p{eq-unf-9}-\p{eq-unf-12}.

The system of equations is reduced to their gauge fixed form of equations \p{xy-1}-\p{xy-3} by taking
\be |{\cal V}_{11}\rangle = c_0^{(3)}|{\cal V}\rangle, \ee
\be |{\cal W}_{i,11}^{j,3}\rangle = \sqrt{2}c_0^{(3)}|{\cal W}_{i}^{j,3}\rangle, \qquad
|{\cal W}_{i,11}^{3,j}\rangle = \sqrt{2}c_0^{(3)}|{\cal W}_{i}^{3,j}\rangle, \qquad
|{\cal W}_{3,11}^{i,j}\rangle = |{\cal W}_{3}^{i,j}\rangle
 \ee
and putting the remaining terms equal to zero.

\section{Expressions for \texorpdfstring{${\cal W}$}{W} and \texorpdfstring{${\cal X}$}{X} vertices for super Yang-Mills-like systems }\label{Appendix D}
In this appendix we present the solutions for
 ${\cal W}$ and ${\cal X}$ vertices for super Yang-Mills-like systems. Analogous solutions
 for supergravity-like systems can be obtained 
 from the present ones by using  the symmetry properties of 
 the defining equations and of the ansatz of the vertex, as
   explained in section \ref{ssugra}. 
   
   The vertex we solve for is defined in equation \p{vertex-1}. In the following we omit the subscript `1' from $Z_{111}$ and $K_{1}^{(i)}$, and we do not write explicitly the index $'3'$ for the oscillator $\alpha_2^{\mu,+}$, since it is present only in the third Fock space.

Omitting the structure constants,
and using the relations
\bea
&&\tilde Q_{B}^{(3)} 
({\cal V})^{ab} =
c^+_2  (p^{(3)} \cdot \gamma)^{ab}
{\cal F}+ \\ \nonumber
&&+c^{(3),+}_1 (\alpha_2^+ \cdot \gamma)^{ab}
((p^{(2)})^2-  (p^{(1)})^2 ) \left (
\frac{\partial {\cal F}}{\partial {\cal Z} } (\alpha^{(1), +}_1 \cdot \alpha^{(2), +}_1)
+ \frac{\partial {\cal F}}{\partial {\cal K}^{(3)} } \right )
\eea
\bea
&&\tilde Q_{F}^{(1)} 
({\cal V})^{ab} = \\ \nonumber
&&=c^{(1),+}_1 (\alpha_2^+ \cdot \gamma)^{ab} ((p^{(3)})^2-  (p^{(2)})^2 ) \left (
\frac{\partial {\cal F}}{\partial {\cal Z} } (\alpha^{(2), +}_1 \cdot \alpha^{(3), +}_1)
+ \frac{\partial {\cal F}}{\partial {\cal K}^{(1)} } \right )
\eea
\bea
&&\tilde Q_{F}^{(2)} 
({\cal V})^{ab} = \\ \nonumber
&&=c^{(2),+}_1 (\alpha^+_2 \cdot \gamma)^{ab}
((p^{(1)})^2-  (p^{(3)})^2 ) \left (
\frac{\partial {\cal F}}{\partial {\cal Z} } (\alpha^{(3), +}_1 \cdot \alpha^{(1), +}_1)
+ \frac{\partial {\cal F}}{\partial {\cal K}^{(2)} } \right )
\eea
  one can solve the equations 
\be \label{xy-1b}
 ({ g}_{0}^{(1)})^a{}_b | {\cal W}_1^{2,3} \rangle^{bc}
 +
 ({g}_{0}^{(2)})^c{}_b | {\cal W}_2^{1,3} \rangle^{ba}
+ 
{\tilde  Q}_{B}^{(3)} 
|{\cal V} \rangle^{ac} =0
\ee
\be \label{xy-2b}
 ({g}_{0}^{(1)})^a{}_b | {\cal W}_1^{3,2} \rangle^{bc}
-
l_0^{(3)}  
|{\cal W}_{3}^{1,2} \rangle^{ac}-
{\tilde Q}_{F}^{(2)} | {\cal V} \rangle^{ac}
=0
\ee
\be \label{xy-3b}
({g}_{0}^{(2)})^a{}_b | {\cal W}_2^{3,1} \rangle^{bc}  
-
l_0^{(3)}  
|{\cal W}_{3}^{2,1} \rangle^{ac}-
{\tilde Q}_{F}^{(1)} | {\cal V} \rangle^{ca}
=0
\ee  
  to get for ${\cal W}$-vertices
\be \label{s-w-ym-1}
({\cal W}_3^{1,2})^{ab}= \,c^{(2),+}_1
 ( \gamma \cdot \alpha_2^+)^{ab}
 \left (
\frac{\partial {\cal F}}{\partial {\cal Z} } (\alpha^{(3), +}_1 \cdot \alpha^{(1), +}_1)
+ \frac{\partial {\cal F}}{\partial {\cal K}^{(2)} } \right )  
\ee
\be
({\cal W}_3^{2,1})^{ab}= - c^{(1),+}_1
  (\gamma \cdot \alpha_2^{+} )^{ab} 
  \left (
\frac{\partial {\cal F}}{\partial {\cal Z} } (\alpha^{(2), +}_1 \cdot \alpha^{(3), +}_1)
+ \frac{\partial {\cal F}}{\partial {\cal K}^{(1)} } \right ) 
\ee
\be
({\cal W}_1^{3,2})^{ab}= \,c^{(2),+}_1
 (p^{(1)} \cdot \gamma)^a{}_c
 (\gamma \cdot \alpha_2^{+} )^{cb}
 \left (
\frac{\partial {\cal F}}{\partial {\cal Z} } (\alpha^{(3), +}_1 \cdot \alpha^{(1), +}_1)
+ \frac{\partial {\cal F}}{\partial {\cal K}^{(2)} } \right ) 
\ee
\be
({\cal W}_2^{3,1})^{ab}=  -c^{(1),+}_1
 (p^{(2)} \cdot \gamma)^a{}_c ( \gamma \cdot \alpha_2^{+})^{cb}
 \left (
\frac{\partial {\cal F}}{\partial {\cal Z} } (\alpha^{(2), +}_1 \cdot \alpha^{(3), +}_1)
+ \frac{\partial {\cal F}}{\partial {\cal K}^{(1)} } \right )  
\ee
\bea \label{s-w-ym-6-1}
({\cal W}_1^{2,3})^{ab}&=& -c^+_2
  C^{ab} {\cal F}+ \\ \nonumber
&+&  c_1^{(3),+}
(p^{(1)} \cdot \gamma)^a{}_c 
(\gamma \cdot \alpha_2^+)^{cb}
\left (
\frac{\partial {\cal F}}{\partial {\cal Z} } (\alpha^{(1), +}_1 \cdot \alpha^{(2), +}_1)
+ \frac{\partial {\cal F}}{\partial {\cal K}^{(3)} } \right )    
\eea
\bea \label{s-w-ym-6}
({\cal W}_2^{1,3})^{ab}&=&- c_2^+ C^{ab} {\cal F}
- \\ \nonumber
&-&c_1^{(3),+}
 (p^{(2)} \cdot \gamma)^a{}_c 
 (\gamma \cdot \alpha_2^+)^{cb}
 \left (
\frac{\partial {\cal F}}{\partial {\cal Z} } (\alpha^{(1), +}_1 \cdot \alpha^{(2), +}_1)
+ \frac{\partial {\cal F}}{\partial {\cal K}^{(3)} } \right )  
\eea
Similarly, from the requirement of preservation of group structure for gauge transformations
\p{inv-w-1}--\p{inv-w-3}
we get
\be \label{inv-w-1b}
 {\tilde Q}_{F}^{(2)} | {\cal W}_1^{2,3} \rangle^{ab}
- \tilde Q_{B}^{(3)} | 
{\cal W}_{1}^{3,2} \rangle^{ab} =
{\tilde Q}_F^{(1)} | {\cal X}_1 \rangle^{ab}
\ee
\be \label{inv-w-2b}
  {
  \tilde Q}_{F}^{(1)} | {\cal W}_2^{1,3} \rangle^{ab}
- \tilde Q_{B}^{(3)} | 
{\cal W}_{2}^{3,1} \rangle^{ab} =
{\tilde Q}_F^{(2)} | {\cal X}_2 \rangle^{ab}
\ee
\be \label{inv-w-3b}
 {\tilde Q}_{F}^{(1)}
| {\cal W}_3^{1,2} \rangle^{ab}
+
{\tilde Q}_{F}^{(2)}  
|{\cal W}_3^{2,1} \rangle^{ba}
=
\tilde Q_{B}^{(3)} 
|{\cal X}_{3} \rangle^{ab}
\ee
Using  the solutions  \p{s-w-ym-1} --  \p{s-w-ym-6}
one can solve for
${\cal X}$-vertices,
\bea \label{x-sol-1}
{\cal X}_1^{ab}&=&  c_1^{(2),+} c_2^+ b_1^{(1),+} C^{ab}
\frac{\partial {\cal F}}{\partial {\cal Z}} (p^{(1)} \cdot \alpha_1^{(3),+}) + \\ \nonumber
&+&  c_1^{(2),+} c_1^{(3),+} b_1^{(1),+}
(p^{(1)}\cdot \gamma)^a{}_c (\gamma \cdot \alpha_2^+)^{cb}
\times \\  \nonumber
&\times&
[-\frac{\partial {\cal F} }{\partial {\cal Z}}   
+\frac{\partial^2 {\cal F}}{\partial {\cal Z}\partial {\cal K}^{(2)}} (p^{(1)} \cdot \alpha_1^{(2),+})
-\frac{\partial^2 {\cal F}}{\partial {\cal Z}\partial {\cal K}^{(3)}}(p^{(1)} \cdot \alpha_1^{(3),+})
+ 
\\ 
\nonumber 
&+&
\frac{\partial^2 {\cal F}}{\partial {\cal Z}^2}
(- (\alpha_1^{(1)+} \cdot \alpha_1^{(2),+})
(p^{(1)} \cdot \alpha_1^{(3),+})
+ (\alpha_1^{(3)+} \cdot \alpha_1^{(1),+})
(p^{(1)} \cdot \alpha_1^{(2),+}) )]
\eea
\bea \label{x-sol-2}
{\cal X}_2^{ab}&=& -c_2^+  c_1^{(1),+}  b_1^{(2),+} C^{ab}
\frac{\partial {\cal F}}{\partial {\cal Z}} (p^{(2)} \cdot \alpha_1^{(3),+}) - \\ \nonumber
&-& c_1^{(3),+} c_1^{(1),+} b_1^{(2),+}
(p^{(2)}\cdot \gamma)^a{}_c (\gamma \cdot \alpha_2^+)^{cb}
\times \\  \nonumber
&\times&
[-\frac{\partial {\cal F} }{\partial {\cal Z}}   
+\frac{\partial^2 {\cal F}}{\partial {\cal Z}\partial {\cal K}^{(3)}} (p^{(2)} \cdot \alpha_1^{(3),+})
-\frac{\partial^2 {\cal F}}{\partial {\cal Z}\partial {\cal K}^{(1)}} (p^{(2)} \cdot \alpha_1^{(1),+}) 
+ 
\\ 
\nonumber 
&+&
\frac{\partial^2 {\cal F}}{\partial {\cal Z}^2}
(- (\alpha_1^{(2)+} \cdot \alpha_1^{(3),+})
(p^{(2)} \cdot \alpha_1^{(1),+})
+ (\alpha_1^{(1)+} \cdot \alpha_1^{(2),+})
(p^{(2)} \cdot \alpha_1^{(3),+}) )]
\eea
\bea \label{x-sol-3}
{\cal X}_3^{ab}&=& c_1^{(1),+} c_1^{(2),+}b_1^{(3)+}
(\gamma \cdot \alpha_2^+)^{ab}
\times \\  \nonumber
&\times&  [ -
\frac{\partial {\cal F} }{\partial {\cal Z}}   
+\frac{\partial^2 {\cal F}}{\partial {\cal Z}\partial {\cal K}^{(1)}} (p^{(3)} \cdot \alpha_1^{(1),+})
-\frac{\partial^2 {\cal F}}{\partial {\cal Z}\partial {\cal K}^{(2)}}(p^{(3)} \cdot \alpha_1^{(2),+})
+ 
\\ 
\nonumber 
&+& 
\frac{\partial^2 {\cal F}}{\partial {\cal Z}^2}
(- (\alpha_1^{(3)+} \cdot \alpha_1^{(1),+})
(p^{(3)} \cdot \alpha_1^{(2),+})
+ (\alpha_1^{(3)+} \cdot \alpha_1^{(2),+})
(p^{(3)} \cdot \alpha_1^{(1),+}) )]
\eea
This completes our treatment of the cubic vertices for the Super Yang-Mills like systems.

\section{Some expressions for linearized gravity} \label{AppendixB}\setcounter{equation}0  
In this section we collect some  expressions for linearized gravity, which we use
for extracting the cubic part from the supergravity Lagrangian.

The metric, the vierbein, the spin connection, and Christoffel symbols are 
\be
g_{{\hat \mu} {\hat \nu}} = \eta_{{\hat \mu} {\hat \nu}} + h_{{\hat \mu} {\hat \nu}}, \quad 
g^{{\hat \mu} {\hat \nu}} = \eta^{{\hat \mu} {\hat \nu}} - h^{{\hat \mu} {\hat \nu}},
\ee
\be
e_{{\hat \mu}}^\mu = \delta_{\hat \mu}^\mu + \frac{1}{2} h_{{\hat \mu}}^\mu,
\quad
e^{{\hat \mu}}_\mu = \delta^{\hat \mu}_\mu - \frac{1}{2} h^{{\hat \mu}}_\mu,
\ee
\be
\omega_{\hat \mu}{}^{\nu \rho} =
 -\frac{1}{2} (\partial^\nu h_{{\hat \mu}}^\rho -
 \partial^\rho h_{{\hat \mu}}^\nu),
\ee
\be
\Gamma^{\hat \rho}_{{\hat \mu} {\hat \nu} }=
\frac{1}{2} \eta^{{\hat \rho} {\hat \lambda}}
(
\partial_{{\hat \mu}} h_{{\hat \lambda} {\hat \nu}}
+
\partial_{{\hat \nu}} h_{{\hat \lambda} {\hat \mu}}
-
\partial_{{\hat \lambda}} h_{{\hat \mu} {\hat \nu}}
)
\ee
The Ricci tensor reads
\be
R_{{\hat \mu} \hat \nu} = -\frac{1}{2}
(\Box h_{{\hat \mu} {\hat \nu}} -
\partial_{{\hat \mu}} \partial_{{\hat \lambda}} h^{{\hat \lambda}}_{{\hat \nu}}-
\partial_{{\hat \nu}} \partial_{{\hat \lambda}} h^{{\hat \lambda}}_{{\hat \mu}}
+
\partial_{{\hat \mu}} \partial_{{\hat\nu}} h_{{\hat \lambda}}^{{\hat \lambda}}
)
\ee
In the equations above the indices are raised and lowered using the flat metric $\eta_{{\hat \mu}{\hat \nu}}$.
The covariant derivative acting on vectors and spin--vectors is defined as follows
\be
\nabla_{\hat \mu} A_{\hat \nu} = \partial_{{\hat \mu}} A_{{\hat \nu}} -
\Gamma^{{\hat \lambda}}_{{\hat \mu} {\hat \nu}} A_{{\hat \lambda}}
\ee
\be
\nabla_{{\hat \mu}} \Psi^a_{{\hat \nu}} = D_{{\hat \mu}} \Psi_{{\hat \nu}}^a
-  \Gamma^{{\hat \lambda}}_{{\hat \mu} {\hat \nu}} \Psi_{{\hat \lambda}}^a, \quad
D_{{\hat \mu}} 
\Psi_{{\hat \nu}}^a =
\partial_{{\hat \mu}} \Psi_{{\hat \nu}}^a + 
\frac{1}{4} \omega_{{\hat \mu}}{}^{\rho \sigma} (\gamma_{\rho \sigma}
\Psi_{{\hat \nu}})^a
\ee

The vertices coupling the fermions to the $B$-field in ten dimensional $N=1$ supergravity mentioned at the end of section \ref{gr+b} are
\be \label{vertex_10dSugraB}
\sqrt2({\bar \Psi_\mu}\gamma^{\mu  \tau \sigma \lambda \nu} \Psi_\nu 
+  {\bar \Psi}^\tau \gamma^{ \sigma  } \Psi^\lambda) \partial_\tau B_{\sigma \lambda} - 2({\bar \Psi}_\mu \gamma^{\tau \sigma \lambda} \gamma^\mu \Xi) \partial_\tau B_{\sigma \lambda}
 \ee


\providecommand{\href}[2]{#2}\begingroup\raggedright\endgroup

\end{document}